\documentclass[twocolumn,showpacs,preprintnumbers,amsmath,amssymb]{revtex4}

\usepackage{graphicx,amsmath,amsfonts,latexsym,graphics,epsfig,subfigure,color,makeidx,hyperref,ulem}
\usepackage{multirow}
\usepackage{array}
\newcommand{\PreserveBackslash}[1]{\let\temp=\\#1\let\\=\temp}
\newcolumntype{C}[1]{>{\PreserveBackslash\centering}p{#1}}
\newcolumntype{R}[1]{>{\PreserveBackslash\raggedleft}p{#1}}
\newcolumntype{L}[1]{>{\PreserveBackslash\raggedright}p{#1}}

\begin{document}
\renewcommand{\baselinestretch}{1.3}
\newcommand\beq{\begin{equation}}
\newcommand\eeq{\end{equation}}
\newcommand\beqn{\begin{eqnarray}}
\newcommand\eeqn{\end{eqnarray}}
\newcommand\nn{\nonumber}
\newcommand\fc{\frac}
\newcommand\lt{\left}
\newcommand\rt{\right}
\newcommand\pt{\partial}
\allowdisplaybreaks

\title{Motion of Spinning Particles around Electrically Charged Black Hole in Eddington-inspired Born-Infeld Gravity}
\author{Ke Yang$^a$\footnote{keyang@swu.edu.cn},
 Bao-Min Gu$^{b,c}$\footnote{gubm@ncu.edu.cn},
Yu-Peng Zhang$^{d,e}$\footnote{zyp@lzu.edu.cn, corresponding author}
}

\affiliation{
$^{a}$School of Physical Science and Technology, Southwest University, Chongqing 400715, China\\
$^{b}$Department of Physics, Nanchang University, Nanchang 330031, China\\
$^{c}$Center for Relativistic Astrophysics and High Energy Physics, Nanchang University, Nanchang 330031, China\\
$^{d}$Lanzhou Center for Theoretical Physics, Key Laboratory of Theoretical Physics of Gansu Province, School of Physical Science and Technology, Lanzhou University, Lanzhou 730000, China\\
$^{e}$ Institute of Theoretical Physics \& Research Center of Gravitation,
Lanzhou University, Lanzhou 730000, China}

\begin{abstract}
A test particle possessing spin angular momentum moves along a non-geodesic path due to an additional spin-curvature force. We study the spinning test particle moving in the vicinity of the electrically charged black hole formation in Eddington-inspired Born-Infeld (EiBI) gravity. Through the numerical analysis of its effective potential and orbits, it is found that the orbital eccentricity reduces as the deviation parameter $\kappa$ increases. {By comparing the orbits for the observed stars around Sagittarius A*, we conclude that the observed orbits with too large radii can not give a stringent constraint with acceptable magnitude. To dig out the potential observation effects of the relations between the orbits and parameter $\kappa$, we mainly focus on the orbits in the vicinity of black hole in this paper.} The parameters of inner most stable circular orbit (ISCO) decrease monotonously with $\kappa$ when the spin angular momentum is small, however they change non-monotonously with $\kappa$ when the spin is large enough. Moreover, the spin dependences of ISCO parameters have similar behavior to that of Reissner-Nordstr\"om (RN) black hole. We analyze the causality of the circular orbits by using the superluminal constraint condition as well. As a result, two new parameter regions may emerge in case of large $\kappa$, where the particle has two stable circular orbits with one subluminal and the other superluminal.
\end{abstract}



\pacs{04.70.-s, 04.50.Kd}









\maketitle



\section{Introduction}

It is widely accepted that general relativity (GR) is an effective infrared  gravitational theory and should be modified in the ultraviolet regime. Therefore, modified gravities are helpful to expand our understanding on gravity and may unveil the corner of the unknown quantum gravity theory \cite{Clifton2012}. One of the long-standing problems suffered by GR is that there are inevitable singularities in cosmology and gravitational collapse \cite{Hawking1970}. These singularities are  usually expected to be regularized in quantum gravity. However, inspired by the well-known Born-Infeld electromagnetic theory \cite{Born1934}, which regularizes the divergent self-energy of the electron in classical dynamics by modifying the behavior of electromagnetic filed at very small scales, Deser and Gibbons proposed a gravitational theory to solve the singularity problem at classical level \cite{Deser1998}. They introduced a square root of a determinant involving the metric and Ricci tensor in the gravitational Lagrangian. However, the theory leads to fourth order equations of motion with ghost-like instability in general. In oder to overcome the problem, Vollick reconsidered the action and worked in a Palatini formalism \cite{Vollick2004}. Since the metric and connection are regarded as independent fields now, the theory gives  second order equations of motion and gets rid of the ghostlike instability.

Furthermore, the square root of a determinant of Ricci tensor constructed by only affine connection could trace back to Eddington's pure affine gravitational theory \cite{Eddington1924}. Thus, by slightly generalizing the original Vollick's action, Ba\~nados and Ferreira showed that the theory may avoid the initial Big Bang singularity of the universe \cite{Banados2010}, which is now dubbed as EiBI gravity. The EiBI gravity is totally equivalent to GR in vacuum. However, when matter is included, the theory recovers GR with cosmological constant in sparse or low curvature regions, but approaches Eddington's theory in dense or high curvature regions. Therefore, the theory has received a considerable attention, and some novel properties may arise in  cosmology \cite{Banados2010,Avelino2012a,Scargill2012,Escamilla-Rivera2012,Yang2013,Cho2013a,Cho2014a,Du2014,Cho2015b,Cho2015}, compact stars \cite{Pani2011,Pani2012,Pani2012a,Harko2013a}, black holes \cite{Olmo2014,Sotani2014,Sotani2015,Jana2015,Wei2015,Avelino2016,Avelino2016a,Jayawiguna2019,Guerrero2020,Guerrero2021}, wormholes \cite{Harko2013b,Shaikh2015,Tamang2015,Olmo2015,Olmo2016}, and topological defects \cite{Liu2012,Fu2014,Avelino2020}. Observational constraints from astrophysics, nuclear physics, gravitational wave data, etc. were studied in Refs.~\cite{Avelino2012,Avelino2012b,Jana2017,Avelino2019,Delhom2020,Jimenez2021}. For a recent review on Born-Infeld inspired gravities see \cite{BeltranJimenez2017,Banerjee:2019nnj,Chakraborty:2020yag,Mukherjee:2017fqz} and the references therein.

The motion of test particles in the vicinity of a black hole is an important source of information on the structure of the background spacetime,  and it is still an effective description for the extreme-mass-ratio-inspiral (EMRI) system. Its trajectory is affected by the mass, charge and angular momentum of the central black hole, and also the additional deviation parameters if the black hole is constructed in modified gravities. The changes in the motion state of the test particles will also be reflected in many aspects and will generate some new observable effects. For example, changes in the corresponding properties of the ISCO, changes in the eccentricity of the bound orbits, and changes in the velocity of the test particles will all induce changes in the gravitational waves generated by the corresponding EMRI system. These changes will also be reflected in the corresponding null geodesics around central black holes and lead to the deformed shadow of a black hole. With the successful photographing of the Black Hole Horizon \cite{ETH2019} and the continuous advancement of the space-borne gravitational wave detectors \cite{lisapaper,Shuichi2017a,Shuichi2017b,taiji,taijisource,Luo2016,Gong:2021gvw}, it has gradually become possible to observe the deviation of the black hole in the modified gravity theory.

A point-like test particle moves along the geodesics. However, if the test particle has internal structures, there will be an additional spin-curvature force, and therefore, the particle will not follow the geodesics any more  \cite{Wald1972,Hanson1974}. By considering  the ``pole-dipole" approximation of the particle, the equations of motion for a spinning test particle moving in a curved background are known as the Mathisson-Papapetrou-Dixon (MPD) equations \cite{Mathisson1937,Papapetrou1951,Corinaldesi1951,Tulczyjew1959,Dixon1964}. The properties of  spinning test particles have been studied in different black hole backgrounds, see Refs.~\cite{Suzuki1998,Han2008,Jefremov2015,Harms2016,LukesGerakopoulos2017,Zhang2017,Mukherjee2018,Zhang2018,Zhang2019b,Antoniou2019,Nucamendi2020,Zhang2020f} for examples, and other space-times \cite{Semerak2015,Mukherjee2018a,Toshmatov2019,Benavides-Gallego:2021lqn}.

In this paper, we are interested in investigating the properties of the spinning test particle in the background of a regular black hole described by EiBI gravity in terms of the bound orbits. Especially, among the bound stable circular orbits of a test particle, there is a marginally stable circular orbit, known as ISCO, which separates the stable circular orbits from the unstable ones that plunge into the central black hole. The ISCO is usually regarded as the inner edge of black hole accretion disk, so it plays an important role in the accretion disk theory \cite{Abramowicz2013}. Therefore, studying the corresponding properties of ISCO will be critical for understanding the accretion disk and the final stage of the gravitational waves from the corresponding EMRI system.

The motion of spinless test particle around the electrically charged black hole in EiBI gravity was studied in Ref.~\cite{Sotani2014}, where they found the radius of ISCO with the specific value of the coupling constant can be smaller than that for the extreme case in GR. In this work, the motion deviation induced by the spin of a test particle around the electrically charged black hole in EiBI gravity is studied and the scheme of the paper is as follows. In Sec.~\ref{Sec_EiBI_Black_Hole}, we briefly review the electrically charged  black hole solution in EiBI gravity. In Sec.~\ref{Sec_EoM}, the equations of motion of spinning test particles is introduced. In Sec.~\ref{Sec_Motion}, we investigate the equatorial motion of the spinning test particles in the EiBI black hole. Finally, brief conclusions are presented.

 \section{Electrically Charged Black Hole Solution in EiBI Gravity}\label{Sec_EiBI_Black_Hole}

The action of the EiBI gravity is given by \cite{Banados2010}
\beqn
S_\text{EiBI}(g,\Gamma,\Psi)&=&\frac{1}{8\pi \kappa}\int{d^4 x\lt[\sqrt{|g_{\mu\nu}+\kappa R_{\mu\nu}(\Gamma)|}-\lambda\sqrt{g} \rt]}\nonumber\\
&&+S_\text{M}(g,\Psi),
\label{EiBI_action}
\eeqn
where $R_{\mu\nu}(\Gamma)$ represents the symmetric part of the Ricci tensor built with the connection $\Gamma$, $S_\text{M}(g,\Psi)$ is the action of matter fields $\Psi$ coupled to the metric field $g$ only, and the dimensionless constant $\lambda$ must be different from zero to avoid meaningless results when matter fields are absent. $\kappa$ is the only deviation parameter of the theory and has inverse dimensions to that of cosmological constant. In this work, we only focus on the theory with respect to a positive $\kappa$, since the theory with the positive $\kappa$ shows superior properties than that with the negative \cite{Pani2011,Avelino2012a,Pani2012a,Yang2013}.

The EiBI action \eqref{EiBI_action} will reduce to the Einstein-Hilbert action with cosmological constant $\Lambda=(\lambda-1)/\kappa$ in the weak field limit $\kappa R\ll g$. By contrast, it will approximate to Eddington's action in the strong field limit $\kappa R\gg g$. Since the theory works in the Palatini formalism, where the metric and the connection are treated as independent variables, the equations of motion are obtained by varying the action (\ref{EiBI_action}) with respect to the metric field $g$ and the connection field $\Gamma$ respectively \cite{Banados2010},
\beqn
\sqrt{q}q^{\mu\nu}&=&\lambda \sqrt{g}g^{\mu\nu}-8\pi\kappa\sqrt{g}T^{\mu\nu},\label{Field_Eq_1}\\
q_{\mu\nu}&=&g_{\mu\nu}+\kappa R_{\mu\nu},\label{Field_Eq_2}
\eeqn
where $q_{\mu\nu}$ is the auxiliary metric compatible to the connection $\Gamma$, i.e., $\Gamma^{\lambda}_{\mu\nu}=\fc{1}{2}q^{\lambda\sigma}(q_{\sigma\mu,\nu}+q_{\sigma\nu,\mu}-q_{\mu\nu,\sigma})$, and $q^{\mu\nu}$ is the inverse of $q_{\mu\nu}$.

The general ansatz of a static spherically symmetric black hole metric is assumed as
\beq
ds^2=-\psi^2(r)f(r)dt^2+\fc{dr^2}{f(r)}+r^2(d\theta^2+\sin^2\theta ~d\phi^2). \label{BH_Metric}
\eeq
Then the corresponding auxiliary metric is assumed as the form
\beq
d\tilde s^2=-G^2(r)F(r)dt^2+\fc{dr^2}{F(r)}+H^2(r)(d\theta^2+\sin^2\theta d\phi^2). \label{Auxiliary_Metric}
\eeq

Since the EiBI gravity is completely equivalent to GR when matter is absent, the standard Schwarzschild-AdS/dS black hole in GR is still hold here in the vacuum, i.e., $\psi=1$, $f(r)=1-2M/r-\Lambda r^2/3$ with cosmological constant $\Lambda=(\lambda-1)/\kappa$. The differences will emerge if one includes the electromagnetic field into the theory,
\beq
\mathcal{L}_\text{M}=-\fc{1}{16\pi}\sqrt{-g}F_{\mu\nu}F^{\mu\nu}.\label{EM_Lagrangian}
\eeq
By varying the matter Lagrangian (\ref{EM_Lagrangian}) with the vector potential, one gets the Maxwell equation
\beq
\nabla_{\mu}F^{\mu\nu}=0.
\eeq
If we only consider the electrostatic field case with the vector potential assumed as $A_\mu=\Phi(r)(dt)_\mu$. Then the field strength $E$ is worked out from the Maxwell equation,
\beq
E=-\pt_r \Phi=\fc{Q}{r^2}\psi,\label{E_Field_Strength}
\eeq
where $Q$ is an integration constant and will be identified as an electric charge eventually.

On the other hand, by substituting the energy-momentum tensor of electromagnetic field
\beq
T^{\mu\nu}=F^{\mu\alpha}{F^{\nu}}_\alpha-\fc{1}{4}g^{\mu\nu}F_{\alpha\beta}F^{\alpha\beta}
\eeq
into the field equation (\ref{Field_Eq_1}), one obtains the formal solutions of variables in auxiliary metric,
\beqn
G&=&  \psi \lt(\lambda-\fc{\kappa Q^2}{r^4}\rt),\label{Exp_G}\\
H&=&r\lt(\lambda+\fc{\kappa Q^2}{r^4}\rt)^\fc{1}{2},\label{Exp_H}\\
F&=&f\lt(\lambda-\fc{\kappa Q^2}{r^4}\rt)^{-1}. \label{Exp_F}
\eeqn
Further, by making use of Eq.~(\ref{Field_Eq_2}), the metric functions are achieved finally \cite{Banados2010} as follows
\beqn
\psi(r) &=&\fc{r^2}{\sqrt{r^4+({\kappa}/{\lambda })Q^2}}, \label{Exp_psi}\\
f(r)&=&-\fc{r\sqrt{\lambda r^4+\kappa Q^2}}{\lambda r^4 -\kappa Q^2}\Bigg[2\sqrt{\lambda}M
       \nn\\
       &&+\int{ \fc{(\Lambda  r^4- r^2+ Q^2)(\lambda  r^4- \kappa Q^2)}{ r^4\sqrt{\lambda  r^4+\kappa Q^2}} d r} \Bigg],\label{Exp_fr_I}
\eeqn
where the corresponding integration constants have been determined properly by checking some asymptotic behaviors of the solution \cite{Sotani2014,Wei2015}. The metric function $f(r)$ can also be expressed as \cite{Wei2015}
\beqn
f(r)&=&\frac{r \sqrt{\kappa  Q^2+\lambda  r^4} }{\kappa  Q^2-\lambda  r^4}\Bigg\{\frac{\left(Q^2-3 r^2+\Lambda  r^4\right) \sqrt{\kappa  Q^2+\lambda  r^4}}{3 r^3}\nn\\
&&-\fc{4}{3}\sqrt{\frac{ i Q^3 }{\sqrt{\kappa\lambda}}}\mathcal{F}\Bigg[ i \sinh ^{-1}\Bigg(r \sqrt{\frac{i \sqrt{\lambda }}{Q\sqrt{\kappa }}}\Bigg),-1\Bigg]\nn\\
&&-\sqrt{\fc{Q^3}{\pi\sqrt{\kappa\lambda}}}\fc{\Gamma[{1}/{4}]^2}{3} +2 \sqrt{\lambda } M \Bigg\},\label{Exp_fr_II}
\eeqn
where $\mathcal{F}$ is the first kind elliptic integral, defined by $\mathcal{F}[\Phi,n]=\int^{\Phi}_0(1-n \sin^2\theta)^{-1/2}d\theta$ with $\pi/2<\Phi<\pi/2$.

\begin{figure}[t]
\begin{center}
\subfigure[$f(r)$]  {\label{f_diff_kappa}
\includegraphics[width=4.1cm,height=3.5cm]{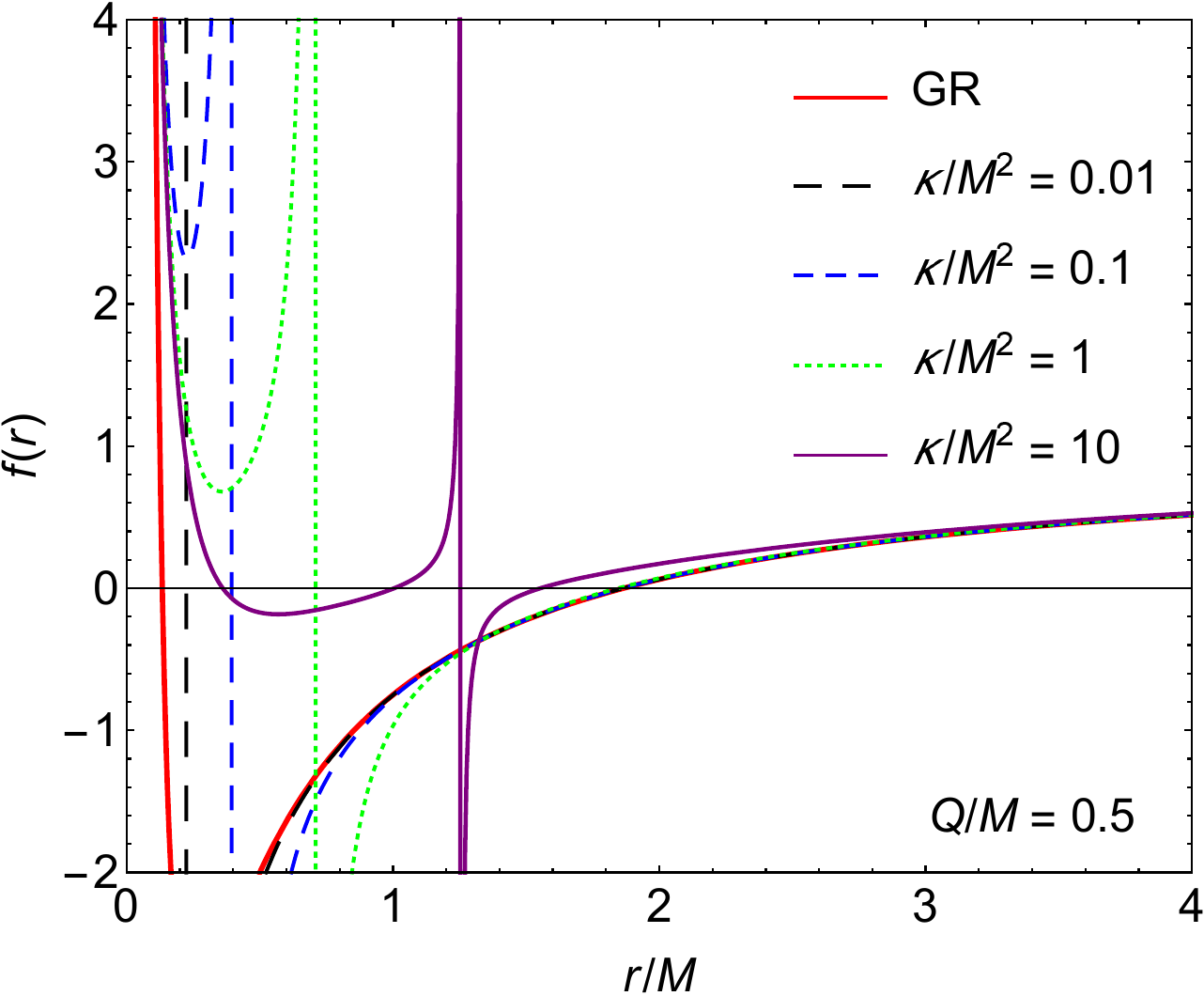}}
\subfigure[Parameter space]  {\label{Parameter_Space}
\includegraphics[width=4.1cm,height=3.5cm]{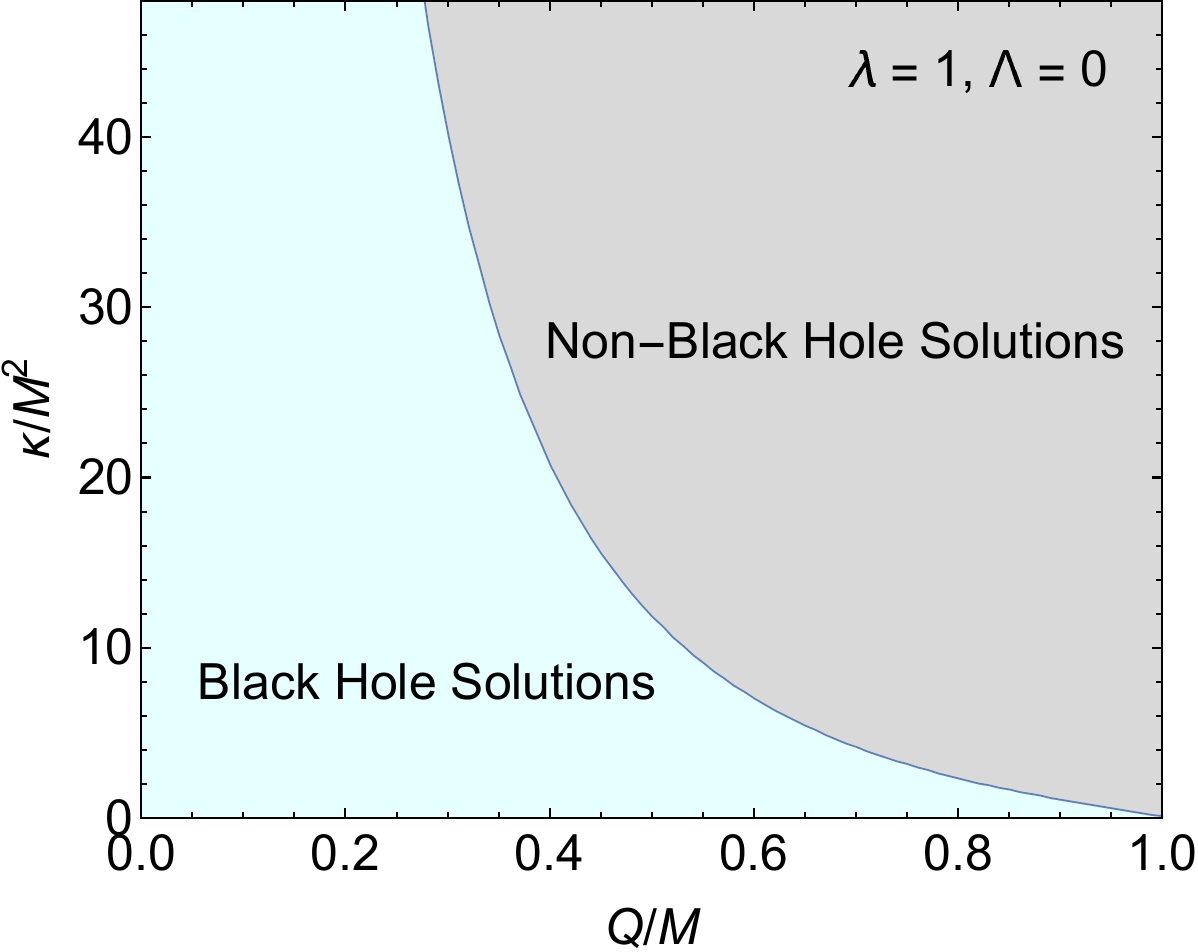}}
\end{center}
\caption{The plots of the metric function $f(r)$ and the parameter space $(Q,\kappa)$ with respect to black holes.}
\end{figure}

The solution will recover the Schwarzschild-AdS/dS black hole solution of GR if the electric charge $Q$ is closed, and will reduce to the RN-AdS/dS solution of GR in the limit of $\kappa\to 0$ just as shown in the Fig.~\ref{f_diff_kappa}. The black hole region in the parameter space $(Q,\kappa)$ is shown in Fig.~\ref{Parameter_Space}, where the region colored in light cyan is relating to the black hole solution, while the region colored in light gray is relating to the solution without horizon.

\section{Equations of Motion for Spinning Test Particles}
\label{Sec_EoM}

A free falling non-spinning particle moves along the geodesics. However, the motion may deviate from the geodesics if the particle has some internal structures \cite{Wald1972,Hanson1974}. If the test particle processes not only single-pole but also dipole terms, its motion is described by the MPD equations \cite{Mathisson1937,Papapetrou1951,Corinaldesi1951,Tulczyjew1959,Dixon1964}
\beqn
\fc{D P^{\mu}}{D \lambda}&=&-\fc{1}{2}R^\mu_{\nu\alpha\beta}u^\nu S^{\alpha\beta}, \label{EoM1}\\
\fc{D S^{\mu\nu}}{D \lambda}&=&P^\mu u^\nu-P^\nu u^\mu, \label{EoM2}
\eeqn
where $R^\mu_{\nu\alpha\beta}$ is the Riemann tensor, $S^{\mu\nu}$ is the spin tensor referring to the particle's internal angular momentum, $P^{\mu}$ is the four-momentum, $u^\mu\equiv dx^\mu/d\tau$ is the four-velocity along the trajectory, $\lambda$ is the affine parameter, and ${D}/{D \lambda}\equiv u^\mu\nabla_\mu$ is the covariant derivative along the trajectory. It is clear that its orbit will differ from the geodesics due to a non-vanishing spin-curvature force on the right side of Eq. \eqref{EoM1}.

Now, by naively counting the degrees of freedom of the variables $S^{\mu\nu}$, $P^\mu$ and $u^\mu$ in the equations of motion (\ref{EoM1}) and (\ref{EoM2}), there are 13 in total, but we have only 10 equations at hand. Thus it is not enough to solve all variables. The underdetermined degrees of freedom are related to the fact that the center of mass of a spinning body is observer-dependent in relativity \cite{Costa2015}. So an additional so called ``spin-supplementary condition" must be given to yield a determinate system. There are different choices of spin-supplementary condition in literatures \cite{Frenkel1926,Mathisson1937,Tulczyjew1959,Dixon1964,Ohashi2003,Kyrian2007}. In this work, we adopt the Tulczyjew spin-supplementary condition \cite{Tulczyjew1959}, given by
\beq
P_{\mu}S^{\mu\nu}=0. \label{Tulczyjew_condition}
\eeq
Combining the Tulczyjew condition with the equations of motion (\ref{EoM1}) and (\ref{EoM2}), one can show that the mass of  the particle $m$  and the spin $s$, defined by
\beqn
m^2&=&-P^\mu P_\mu,\label{mass}\\
s^2&=&\frac{1}{2}S^{\mu\nu}S_{\mu\nu},\label{spin}
\eeqn
are conserved constants of motion.

Further, the equations of motion (\ref{EoM1}) and (\ref{EoM2}) and Tulczyjew condition (\ref{Tulczyjew_condition}) yield the relation
\beq
u^\mu+ \frac{P^\nu u_\nu}{m^2}P^\mu =\fc{S^{\mu\nu}S^{\alpha\beta}R_{\nu\rho\alpha\beta}u^\rho } {2m^2}.
\eeq
It implies that the four-velocity $u^\mu$ is generally not parallel to the four-momentum $P^\mu$ under Tulczyjew condition. This can be more clearly observed if one normalizes the four-velocity as $u_\mu v^\mu=-1$ with $v^\mu=P^\mu/m$ the normalized momentum, then the relation is rewritten as $u^\mu- v^\mu =\fc{S^{\mu\nu}S^{\alpha\beta}R_{\nu\rho\alpha\beta}u^\rho } {2m^2}$.

The four-momentum is always timelike along the trajectory as is shown in Eq.~(\ref{mass}). However, since $u^\mu u_\mu$ is not a constant of motion generally, the four-velocity $u^\mu$ may transform from timelike to spacelike, then the orbit becomes unphysical \cite{Hojmanthesis,Hojman2013}. In order to make sure that the motion of spinning test particles is physical, the superluminal constraint should be adopted \cite{Zhang2017}, i.e.,
\beq
\fc{u^\mu u_\mu}{(u^t)^2}=g_{tt}+g_{rr} \dot r ^2+g_{\phi\phi} \dot \phi ^2+2g_{t\phi}\dot\phi<0,
\label{Superluminal_Constraint}
\eeq
where the dot denotes the derivative with respect to the coordinate time $t$. Furthermore, this superluminal problem of spinning particles with ``pole-dipole" approximation may be avoided by considering a non-minimal spin-gravity interaction through a gravimagnetic moment \cite{Deriglazov2017}.

\section{Motion of Spinning Test Particles in Electrically Charged Black Hole in EiBI Gravity }
\label{Sec_Motion}

\subsection{Equatorial Motion}

We consider the motion of spinning test particles in the background of the black hole \eqref{BH_Metric} in EiBI gravity. Here, we focus on their equatorial motion with the spin-aligned or anti-aligned orbits, i.e., the four-momentum and spin tensor satisfy $P^\theta=0$ and $S^{\mu\theta}=0$. With the Tulczyjew condition (\ref{Tulczyjew_condition}), the non-vanishing components of spin tensor read
\beqn
S^{rt}=-\frac{P_\phi}{P_t}S^{r\phi}, \quad
S^{\phi t}=\frac{P_r}{P_t}S^{r\phi}.
\label{Srt&Sphit}
\eeqn
Inserting above relations into \eqref{spin} yields
\beq
S^{r\phi}=\pm\frac{s}{r}\frac{P_t}{m\psi}.
\eeq
Since the particle is restricted on the equatorial plane, the only non-vanishing component of the spin vector, $S^\mu=-\frac{1}{2}\epsilon^\mu_{\nu\alpha\beta}u^{\nu}S^{\alpha\beta}$ with $\epsilon_{\mu\nu\alpha\beta}$ the completely antisymmetric tensor, is $S^{\theta}$, namely, the spin direction is perpendicular to the equatorial place. Thus the overall signs of $S^{\mu\nu}$ represents the spin-aligned or anti-aligned orbit. Without loss of generality, we choose the minus sign in $S^{r\phi}$, so a positive $s$ represents the spin-aligned case and a negative to anti-aligned. Now all the non-vanishing components of spin tensor are rewritten as
\beqn
S^{r\phi} &=&-S^{\phi r} =-\frac{s}{r}\frac{P_t}{m \psi},\label{ex_srphi}\\
S^{\phi t}&=&-S^{t\phi}=-\frac{s}{r}\frac{P_r}{m \psi},\label{ex_sphit}\\
S^{tr}&=&-S^{rt}=-\frac{s}{r}\frac{P_\phi}{m \psi}.\label{ex_str}
\eeqn

Furthermore, if the background geometry possesses some symmetry, another quantity,
\beq
\mathcal{C}=\mathcal{K}^\mu P_\mu-\frac{1}{2}S^{\mu\nu}\mathcal{K}_{\mu;\nu},
\eeq
is also a constant of motion \cite{Hojman1977}, where $\mathcal{K}^\mu$ is the Killing vector associated with the symmetry and satisfies the Killing equation $\mathcal{K}_{\mu;\nu}+\mathcal{K}_{\nu;\mu}=0$. According to the static spherically symmetric spacetime \eqref{BH_Metric}, the geometry possesses a timelike Killing vector $\xi^\mu=(\pt_t)^\mu$ and a spacelike Killing vector $\eta^\mu=(\pt_\phi)^\mu$. Therefore, there are another two important conserved quantities, i.e., the energy $e$ and the total angular momentum $j$ \cite{Hojman1977,Zhang2016a},
\beqn
e&=&-\xi^\mu P_\mu+\frac{1}{2}S^{\mu\nu}\xi_{\mu;\nu}=-P_t-\fc{1}{2}\fc{s}{r}\fc{P_\phi}{m\psi}g'_{tt},\label{CQ_Energy}\\
j&=&\eta^\mu P_\mu -\fc{1}{2}S^{\mu\nu}\eta_{\mu;\nu}=P_\phi-\fc{1}{2}\fc{s}{r}\fc{P_t}{m\psi}g'_{\phi\phi},\label{CQ_angular_momentum}
\eeqn
where the prime denotes the derivative with respect to the radius $r$.
The non-vanishing components of the momentum can be worked out from above two equations plus \eqref{BH_Metric} and \eqref{mass}, namely,
\beqn
P_t&=&\frac{ m M^2 \bar{s}\bar{j} \left(\psi f'+2\psi' f \right)-2m \bar{e}  r}{2 r-\bar{s}^2M^2  f'-2M^2  \bar{s}^2 f \psi'/\psi },\\
P_\phi &=& \frac{2 m M r (\bar{j}-\bar{e} \bar{s}/\psi)}{2 r- M^2 \bar{s}^2  f' -2  M^2 \bar{s}^2 f \psi'/\psi },\\
(P^r)^2&=&\frac{P_t^2}{\psi^2}-\lt(m^2+\frac{P_\phi ^2}{r^2}\rt)f,
\eeqn
where we have rescaled the parameters of energy, total angular momentum and spin to be dimensionless via $\bar{e}=\fc{e}{m}$, $\bar{j}=\fc{j}{mM}$ and $\bar{s}=\fc{s}{mM}$, respectively.

From the equation of motion \eqref{EoM2}, we have
\beqn
\fc{D S^{tr}}{D\lambda}&=&P^t u^r-P^r,\\
\fc{D S^{t\phi}}{D\lambda}&=&P^t u^\phi-P^\phi.
\eeqn
Then inserting Eqs. \eqref{ex_sphit} and \eqref{ex_str} into above equations, one gets
\beqn
P^t u^r-P^r \!&\!=\!&\!-\fc{M\bar{s}}{r\psi}\fc{DP_\phi}{D\lambda}+\fc{M\bar{s}P_\phi}{r \psi}\lt(\fc{1}{r}+\fc{\psi'}{\psi} \rt)  u^r,\label{RealtionI}\\
P^t u^\phi-P^\phi \!&\!=\!&\!\fc{M\bar{s}}{r\psi}\fc{DP_r}{D\lambda}-\fc{M\bar{s}P_r}{ r\psi}\lt(\fc{1}{r}+\fc{\psi'}{\psi} \rt) u^r.\label{RealtionII}
\eeqn
Further, using the equation of motion \eqref{EoM1}, one obtains that
\beqn
\!\!\!\!\!\!\! P^t u^r \!-\! P^r \!&\!\!=\!\!&\!\fc{M\bar{s}}{2 r\psi}R_{\phi\nu\alpha\beta}u^\nu S^{\alpha\beta}\!+\!\fc{M\bar{s}P_\phi}{r \psi}\lt(\fc{1}{r} \!+\! \fc{\psi'}{\psi} \rt)\!  u^r,\\
\!\!\!\!\!\!\! P^\phi \!-\!  P^t u^\phi  \!&\!\!=\!\!&\! \fc{M\bar{s}}{2 r\psi}R_{r\nu\alpha\beta}u^\nu S^{\alpha\beta} \!+\! \fc{M\bar{s}P_r}{ r\psi}\lt(\fc{1}{r}\!+\!\fc{\psi'}{\psi} \rt)\! u^r.
\eeqn
So the four-velocity can be worked out algebraically, yielding
\beqn
u^r&=&\fc{c_1}{a_1},\label{ur}\\
u^\phi&=&\fc{a_1 c_2-a_2 c_1}{a_1 b_2 },\label{uphi}
\eeqn
where the variables $a_1$, $b_1$, $c_1$, $a_2$, $b_2$ and $c_2$ are defined as follows,
\beqn
a_1&=&P^t-\frac{M \bar{s}}{r \psi } \left[R_{\phi rr\phi } S^{r\phi }+\left(\frac{1}{r}+\frac{\psi '}{\psi }\right)P_{\phi } \right],\\
c_1&=&P^r+\fc{M\bar{s}}{r\psi}R_{\phi t t\phi}S^{t\phi},\label{Para_c1}\\
a_2&=& \frac{M \bar{s} }{r \psi } \lt(\frac{1}{r}+\frac{\psi '}{\psi }\rt)P_r, \\
b_2&=&P^t+\fc{M\bar{s}}{r\psi}R_{r\phi r\phi}S^{r\phi},\\
c_2&=&P^\phi-\fc{M\bar{s}}{r\psi}R_{r t tr}S^{tr}.
\eeqn

From Eqs.~\eqref{ex_sphit}, \eqref{ur} and \eqref{Para_c1}, it is clear that $u^r \propto P_r $. Therefore, although the four-velocity $u^\mu$ is generally not parallel to the four-momentum $P^\mu$ under Tulczyjew condition, the radial velocity $u^r$ is always parallel to the radial momentum $P^r$, and this is helpful to derive the effective potential of spinning test particles later.

Moreover, in order to simplify the calculation, instead of normalizing the four-velocity as  $u_\mu v^\mu=-1$, here we choose to fix $u^t=1$ by setting the affine parameter $\lambda$ as the coordinate time $t$, due to the fact that the trajectories of test particles are independent of the affine parameter. Therefore, the velocities $u^r=\dot r, u^\phi=\dot\phi$.

\subsection{Orbits}

\begin{figure}[t]
\begin{center}
\subfigure[$V_{\text{eff}}$]  {\label{V_diff_Gt}
\includegraphics[width=4.1cm,height=3.5cm]{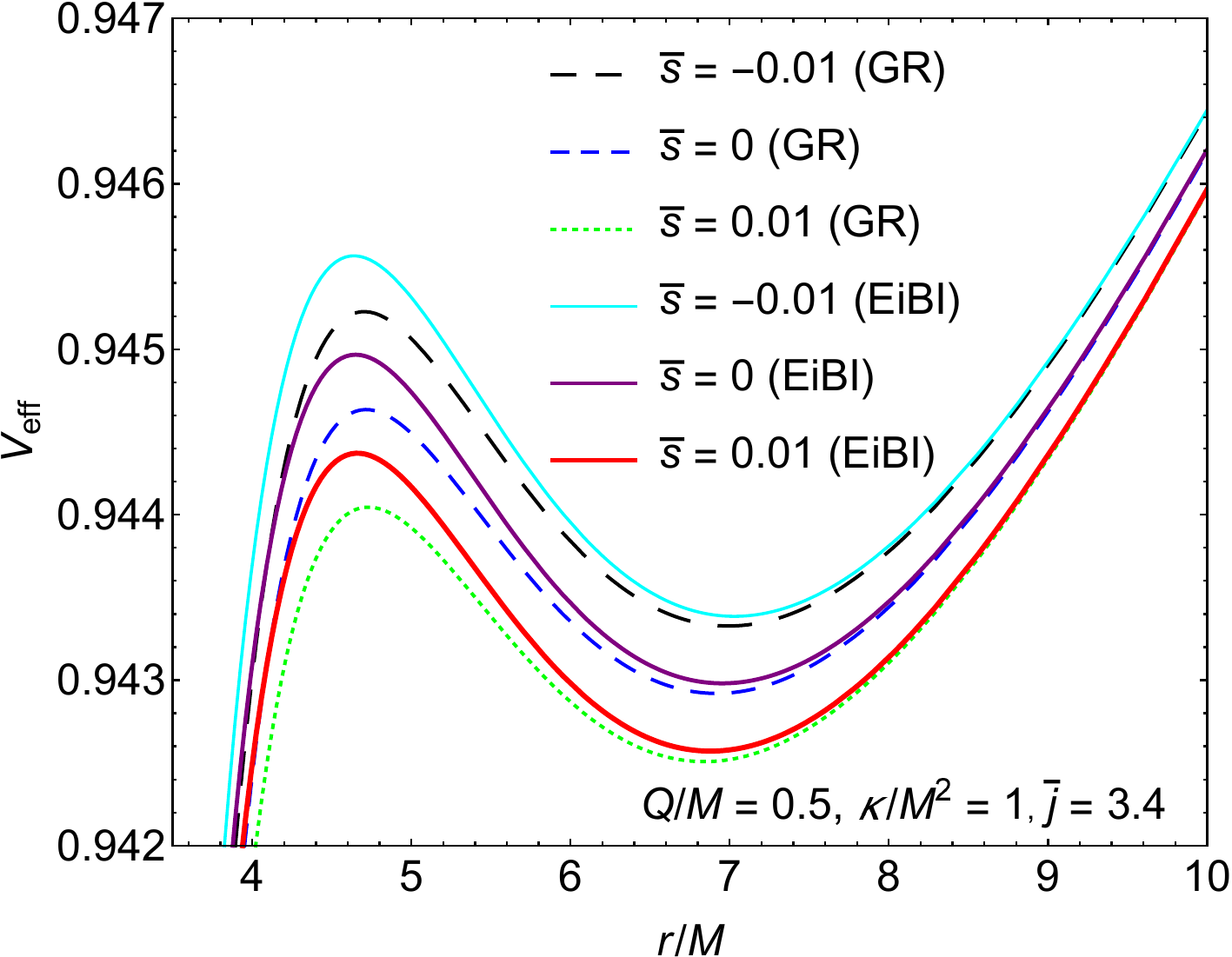}}
\subfigure[$V_{\text{eff}}$]  {\label{V_diff_kappa}
\includegraphics[width=4.1cm,height=3.5cm]{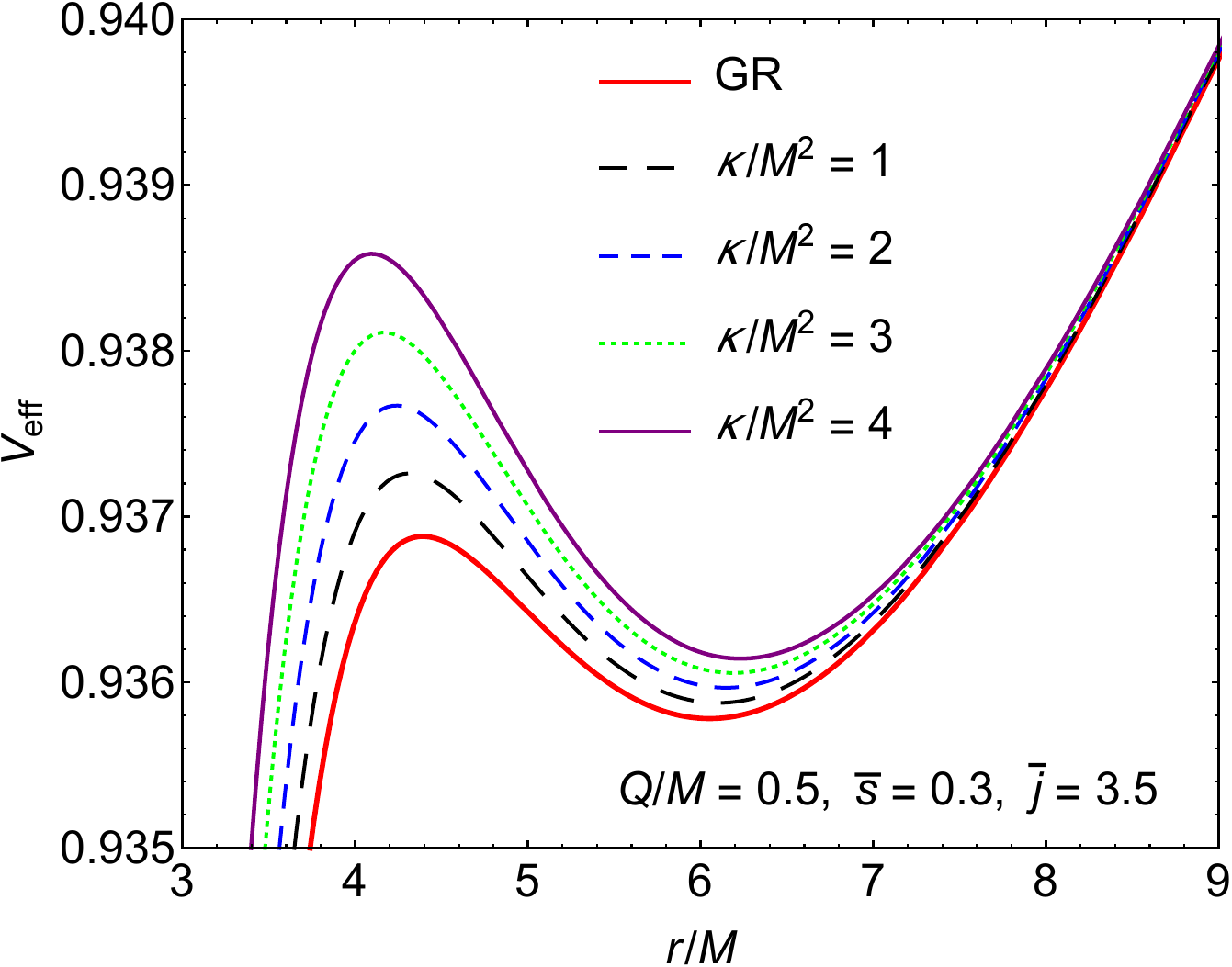}}\\
\subfigure[$V_{\text{eff}}$]  {\label{V_diff_s}
\includegraphics[width=4.1cm,height=3.5cm]{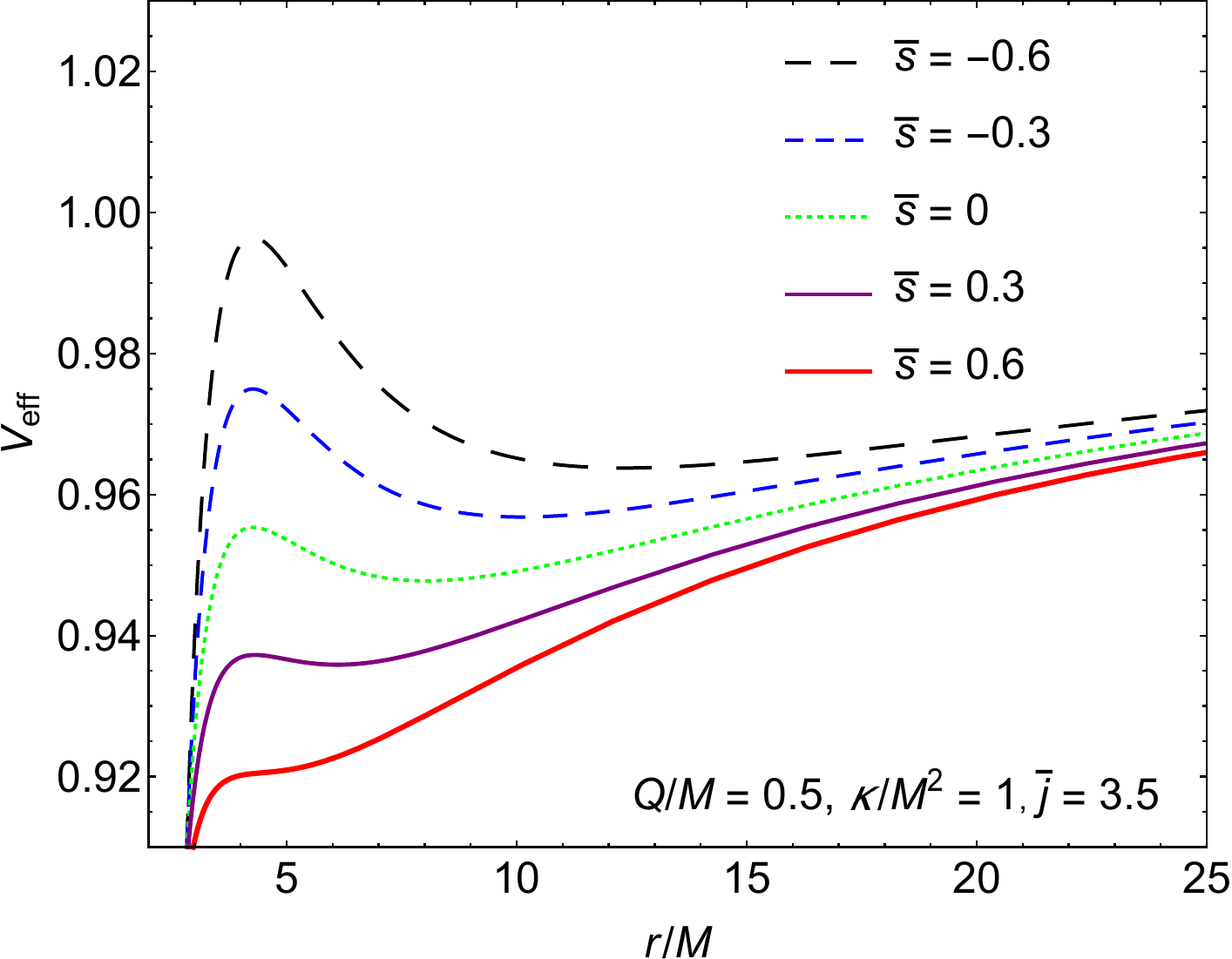}}
\subfigure[$V_{\text{eff}}$]  {\label{V_diff_j}
\includegraphics[width=4.1cm,height=3.5cm]{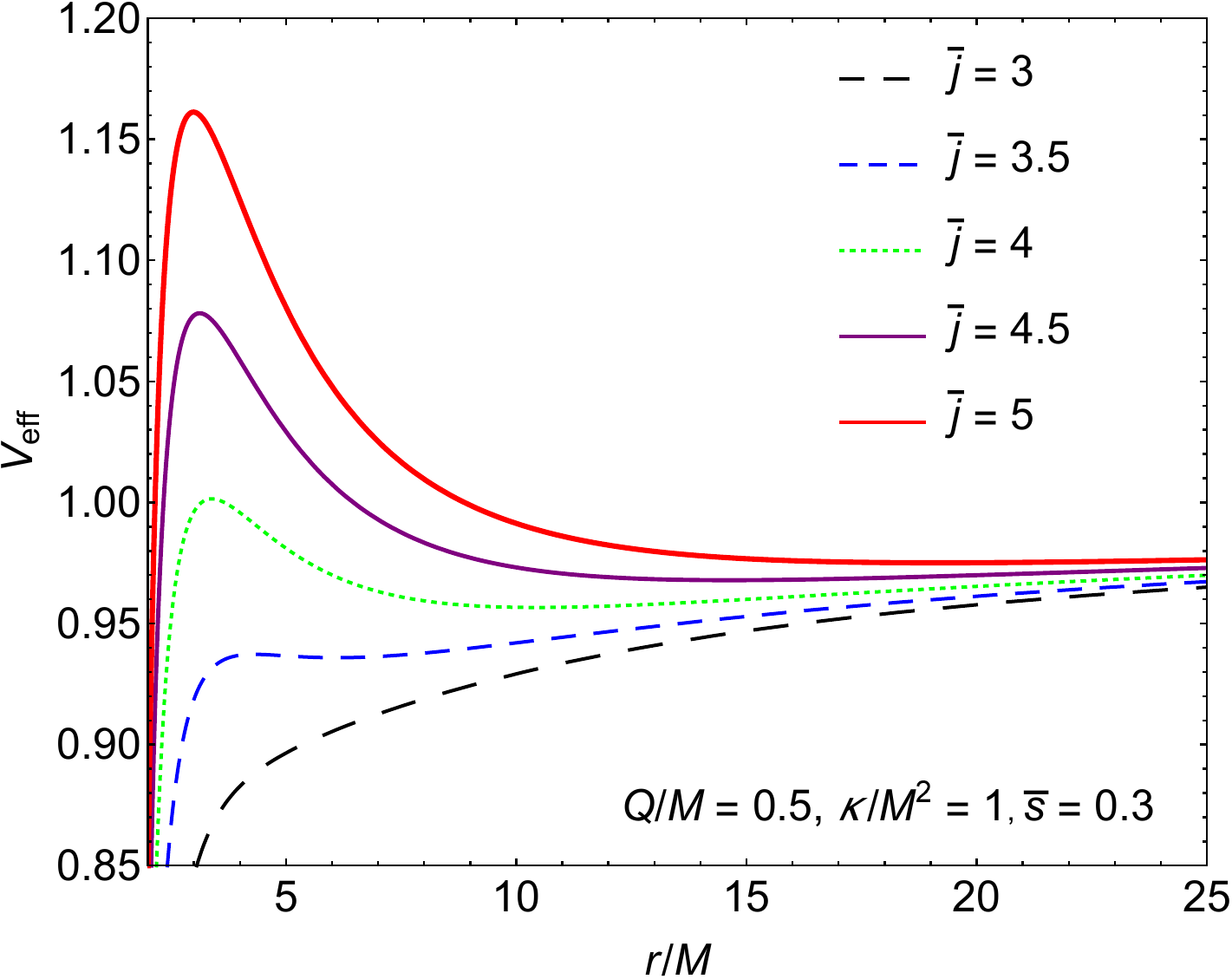}}
\end{center}
\caption{The shapes of the effective potential $V_{\text{eff}}$ for different parameters.}
\label{Fig_Eff_Potential}
\end{figure}

\begin{figure*}[t]
\begin{center}
\subfigure[GR]  {\label{}
\includegraphics[width=3.8cm,height=3cm]{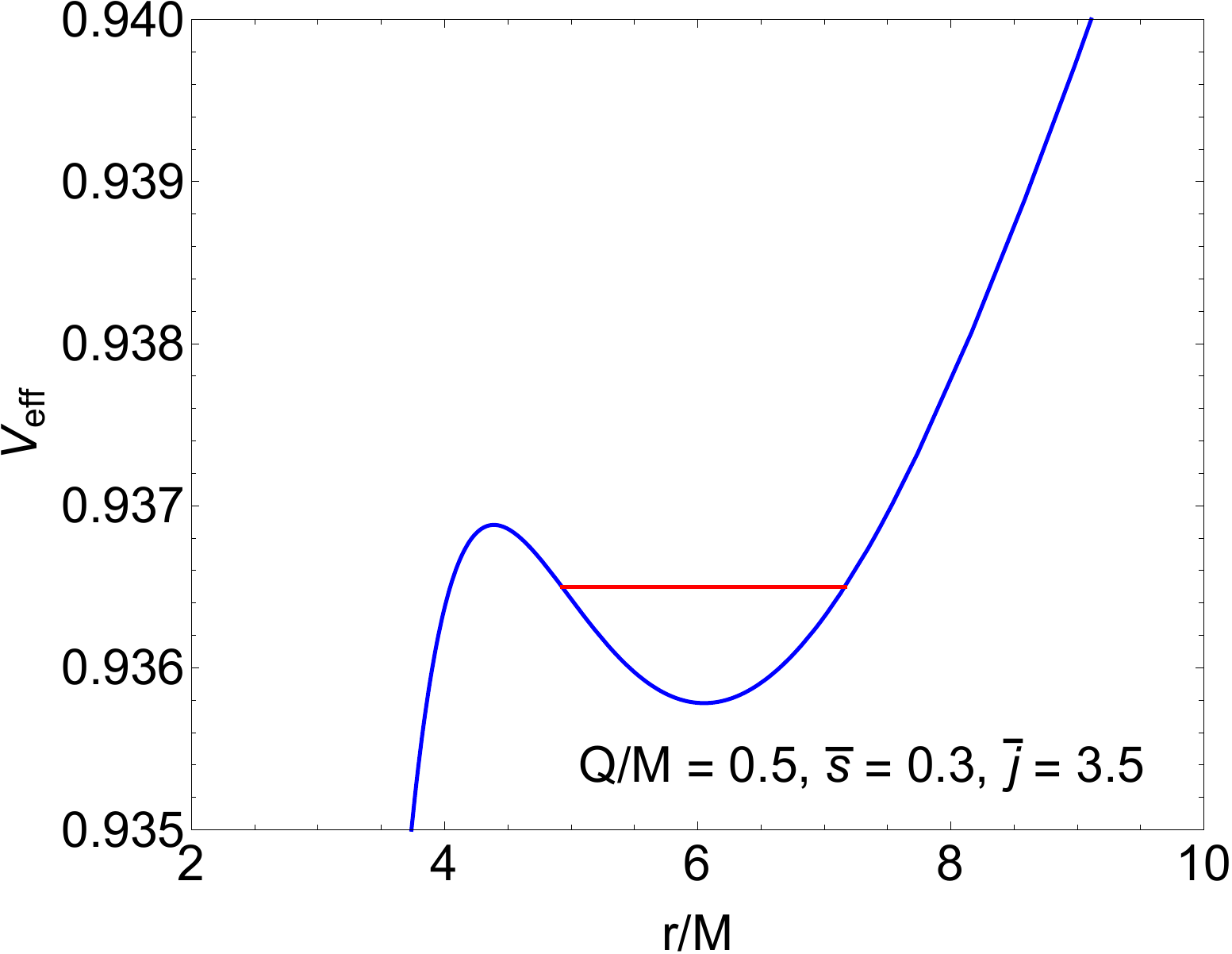}}
\subfigure[$\kappa/M^2=0.1$]  {\label{}
\includegraphics[width=3.8cm,height=3cm]{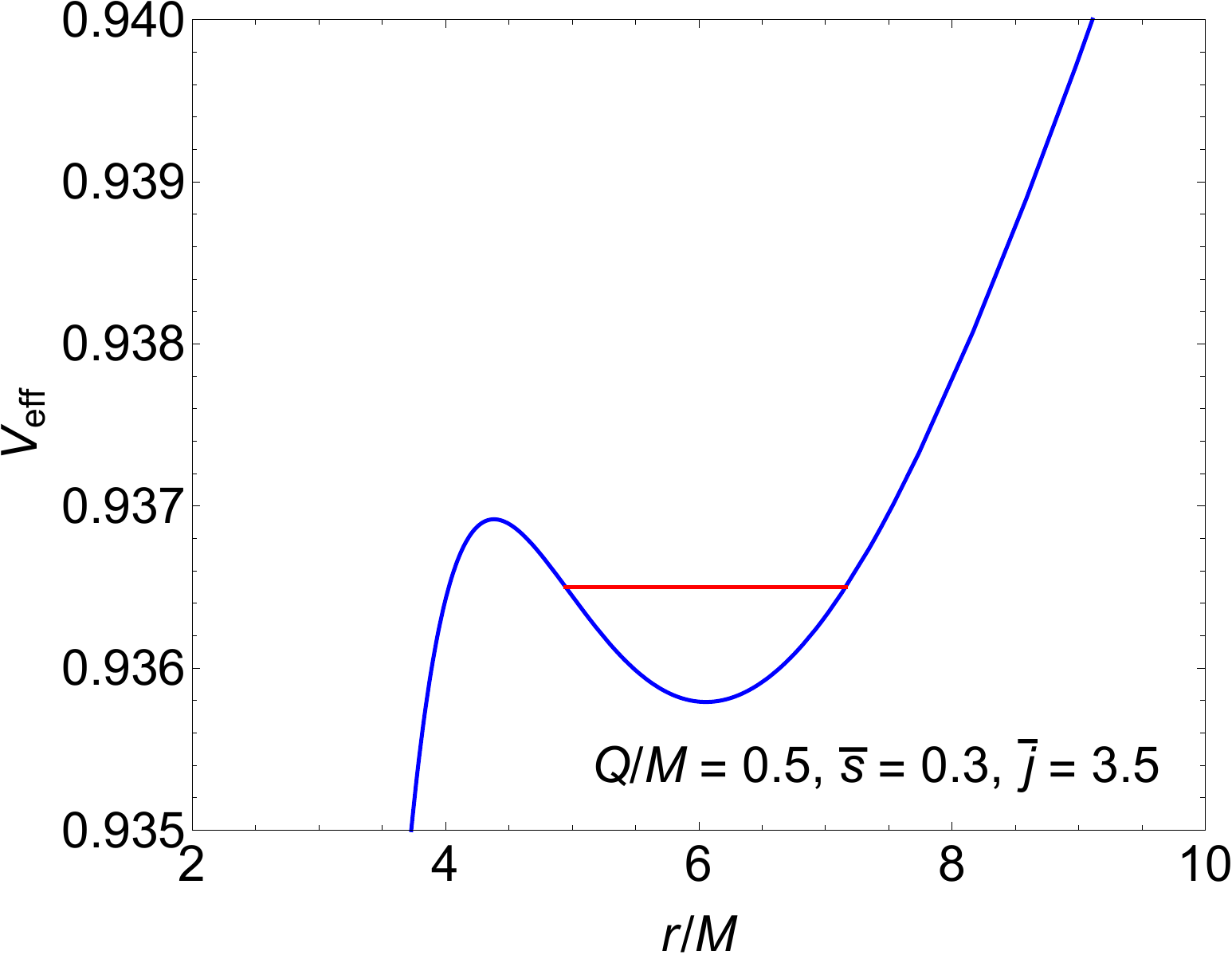}}
\subfigure[$\kappa/M^2=1$]  {\label{}
\includegraphics[width=3.8cm,height=3cm]{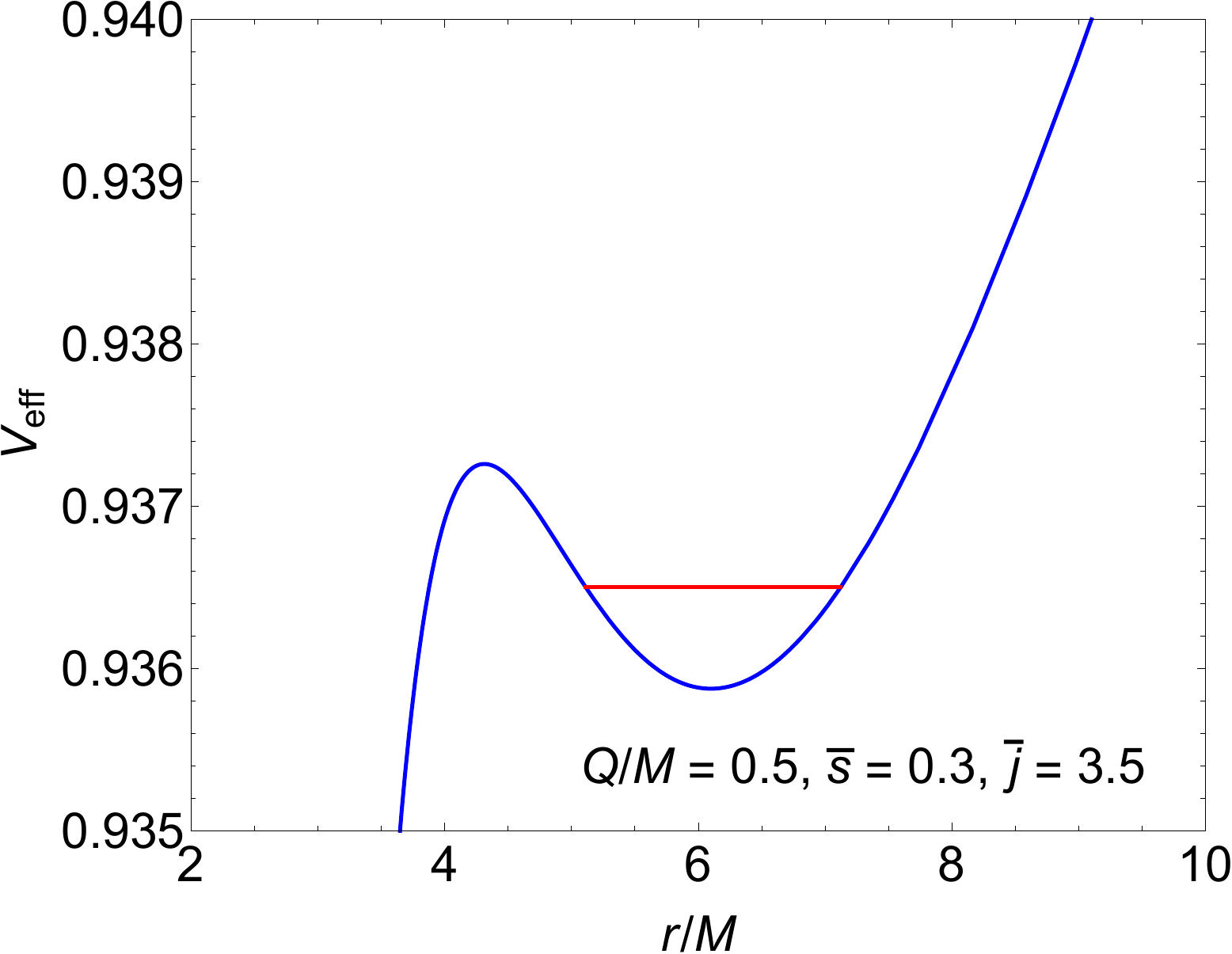}}
\subfigure[$\kappa/M^2=5$]  {\label{}
\includegraphics[width=3.8cm,height=3cm]{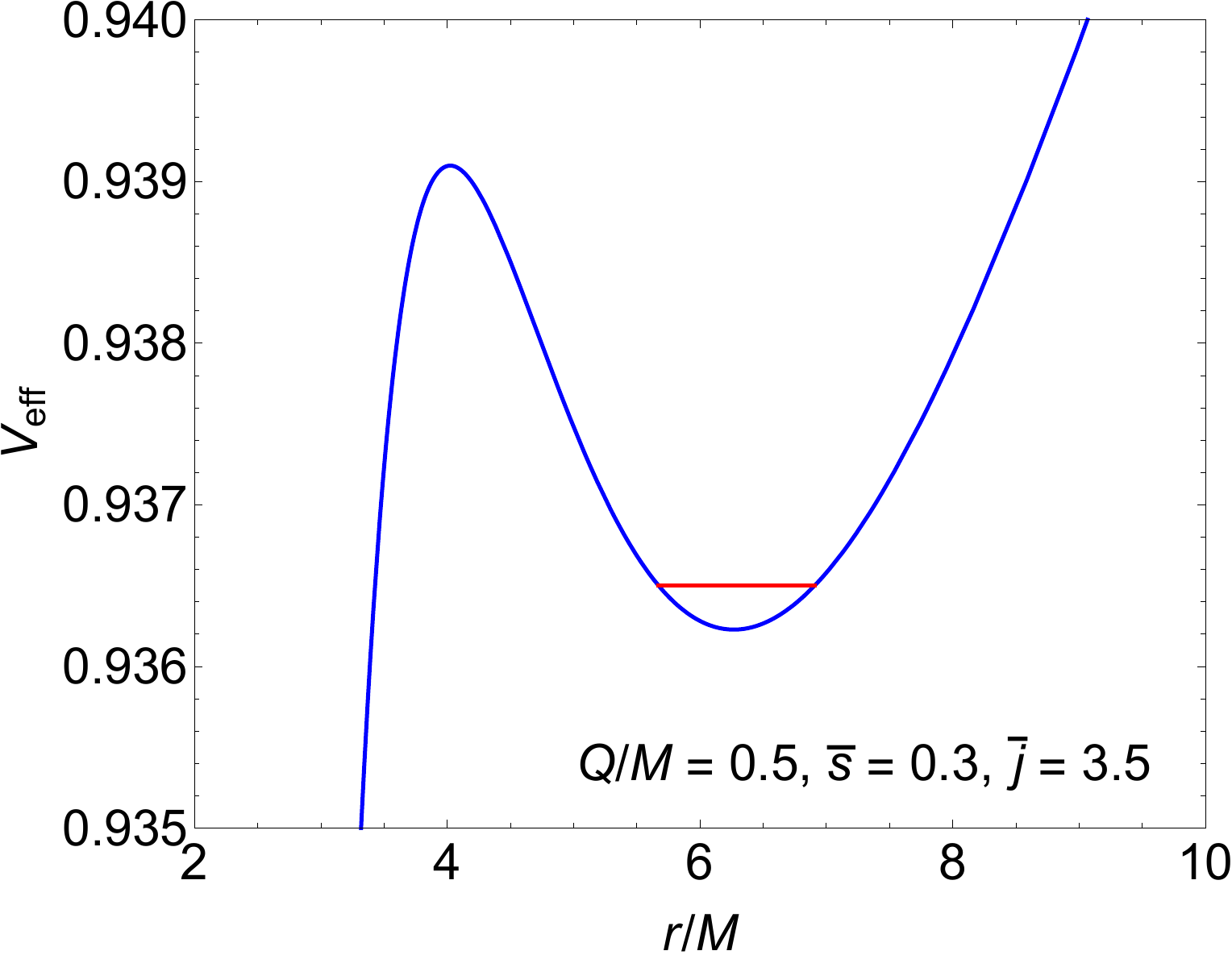}}\\
\subfigure[GR]  {\label{}
\includegraphics[width=3.8cm,height=3.8cm]{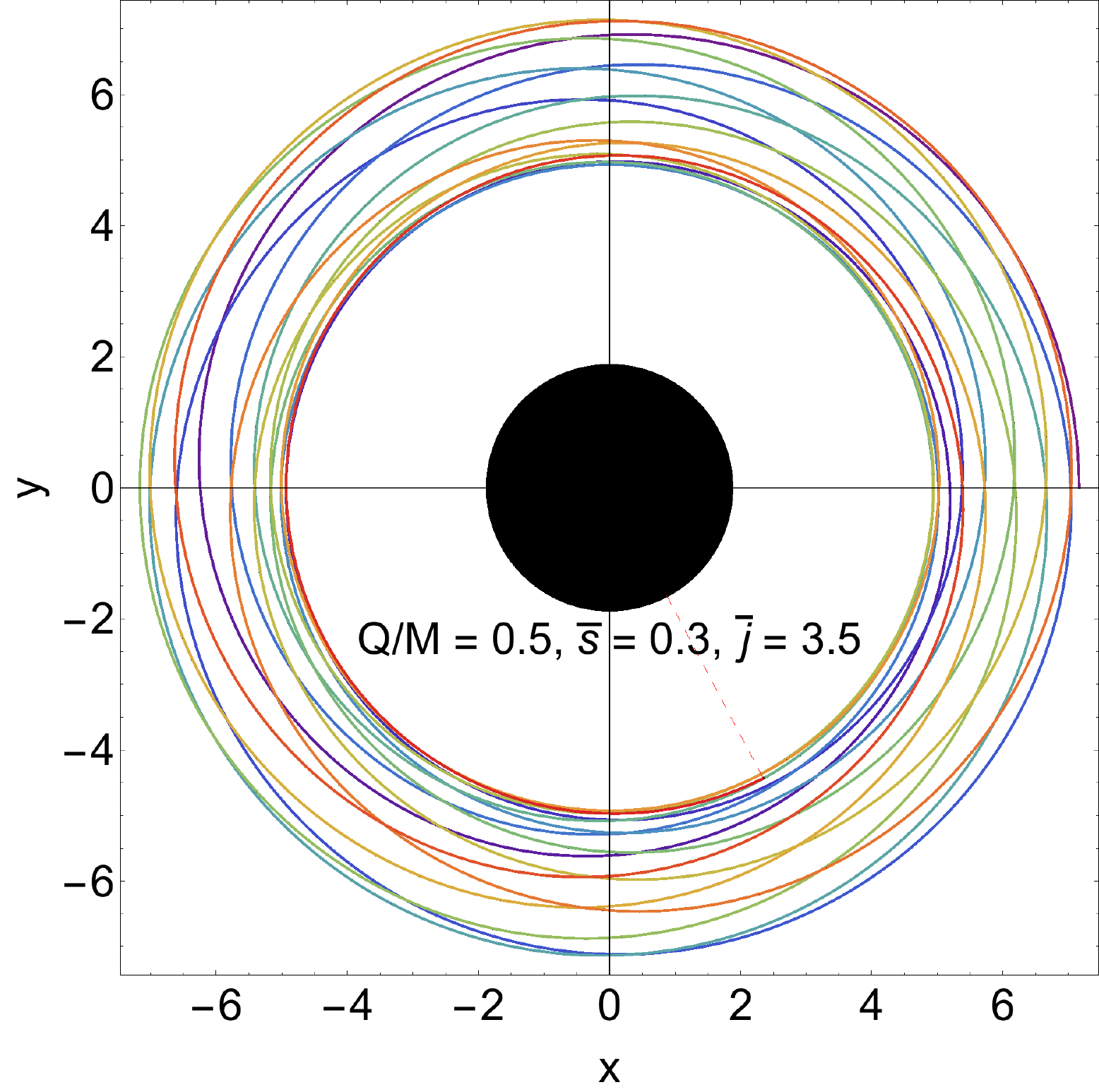}}
\subfigure[$ \kappa/M^2=0.1$]  {\label{}
\includegraphics[width=3.8cm,height=3.8cm]{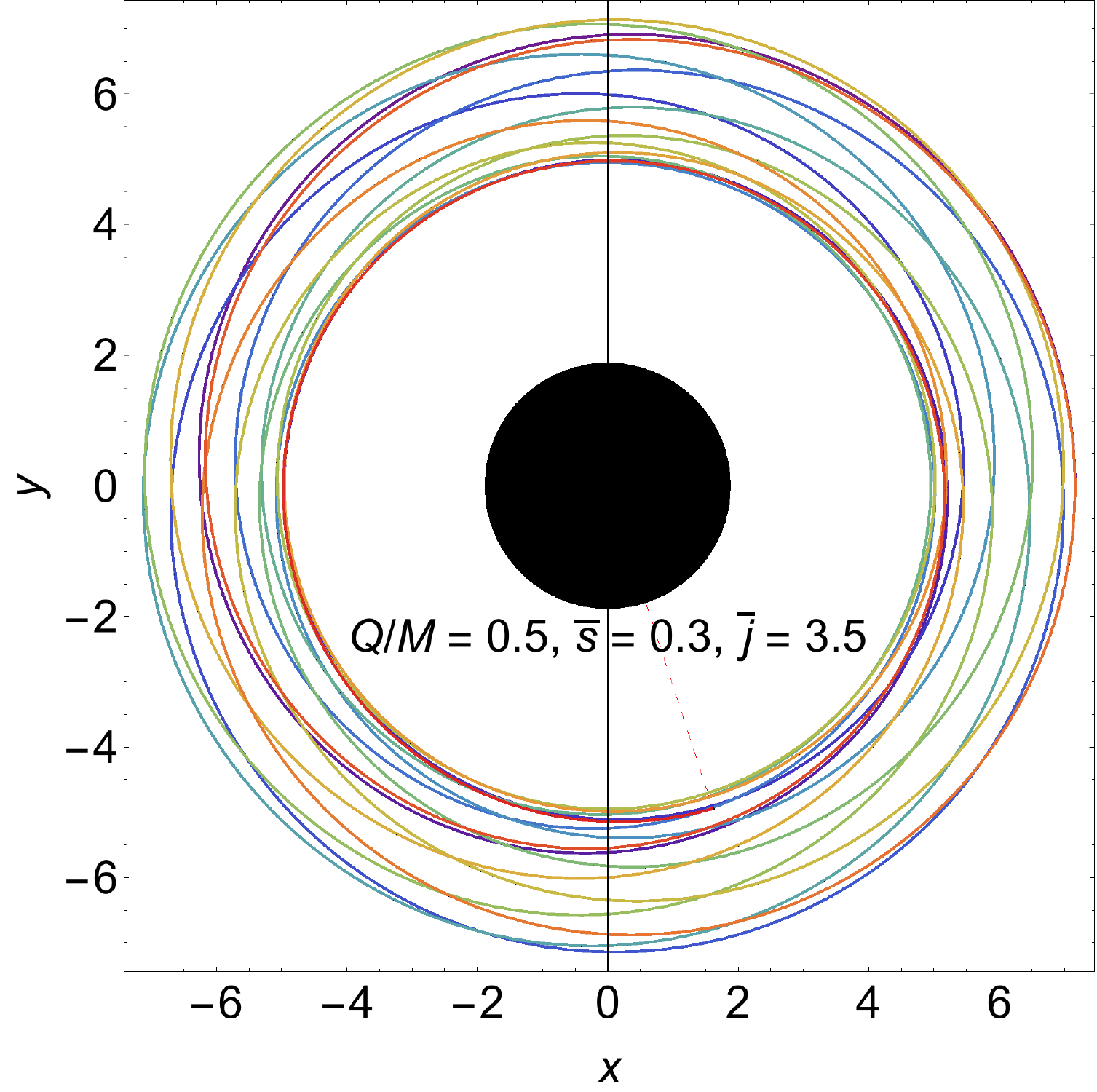}}
\subfigure[ $\kappa/M^2=1$]  {\label{}
\includegraphics[width=3.8cm,height=3.8cm]{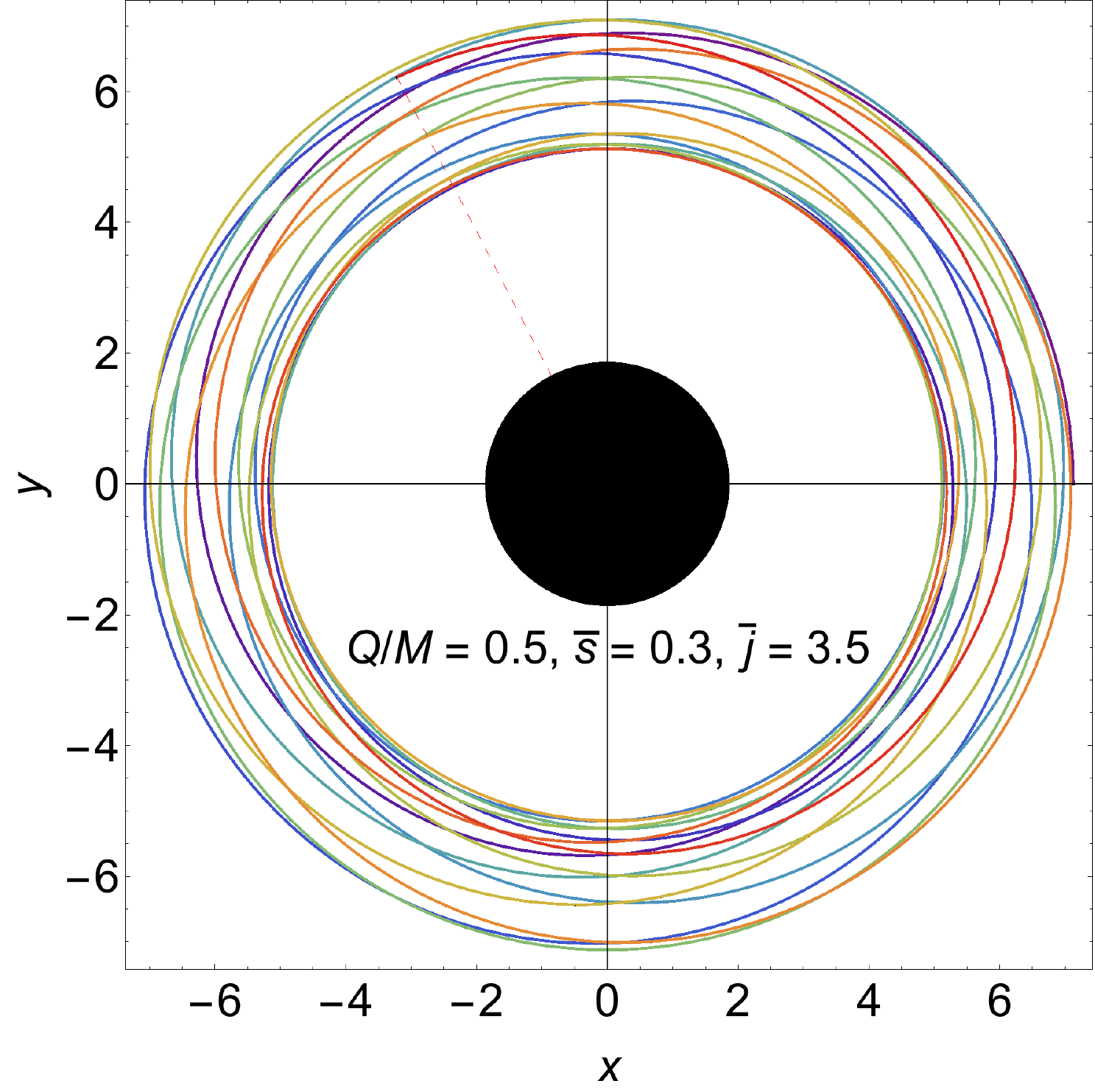}}
\subfigure[ $\kappa/M^2=5$]  {\label{}
\includegraphics[width=3.8cm,height=3.8cm]{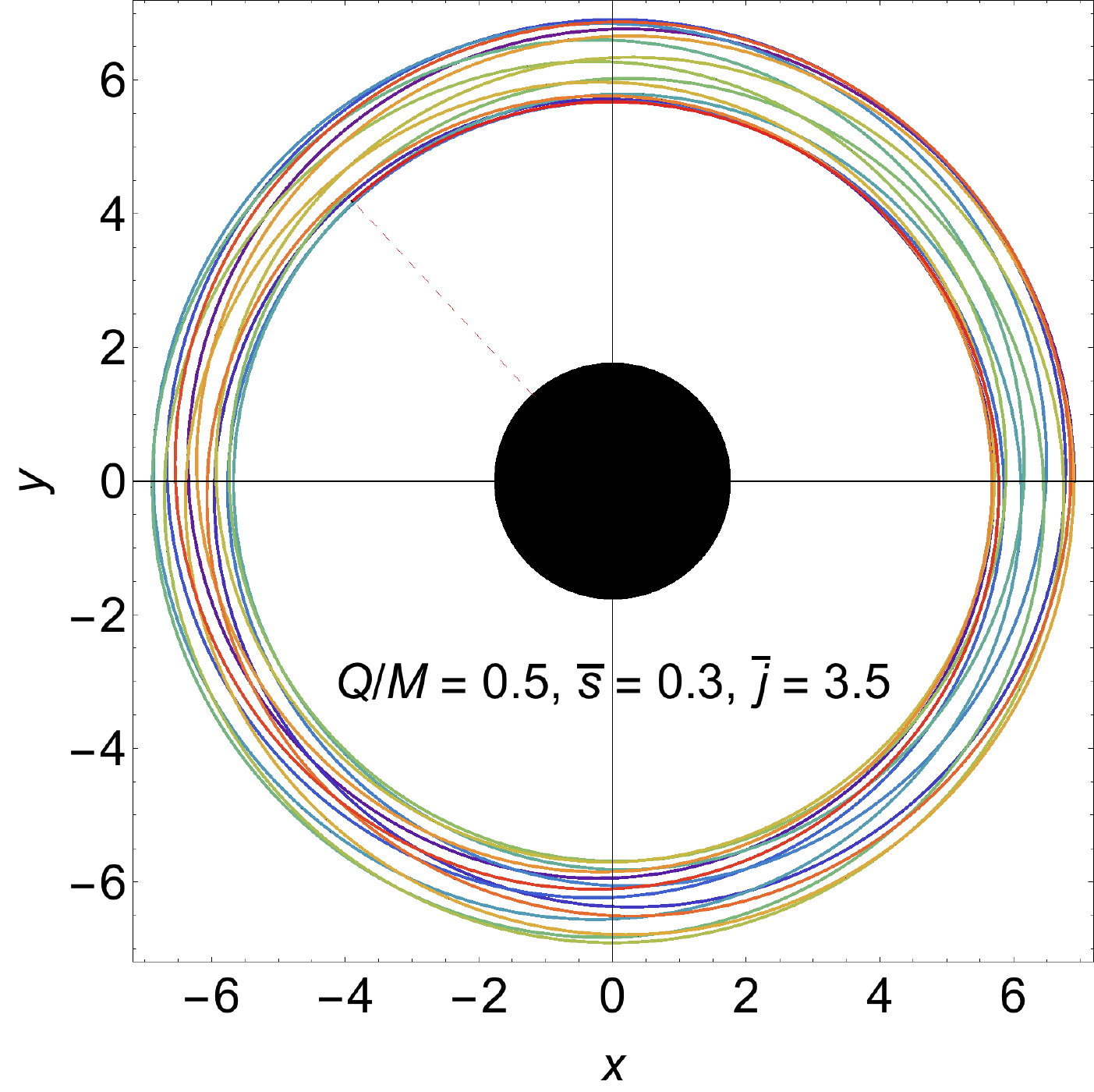}}
\end{center}
\caption{The shapes of the effective potential $V_{\text{eff}}$ for different $\kappa$ (3a-3d), where the horizontal red lines in potential indicate the particle energy $\bar e=0.9365$. The corresponding orbits with respect to the energy $\bar e=0.9365$ are plotted in (3e-3f), where the dashed red lines indicate the terminal points of the trajectories at the end of the time interval, and the black discs represent the inner regions of black holes.}
\label{Fig_Orbit}
\end{figure*}

The effective potential method is an effective way to investigate the motion of a classical test particle in a central field. Once we get the effective potential of the spinning test particle in a black hole background, its properties of motion can be explored from the effective potential \cite{Jefremov2015}. Since the radial velocity $u^r$ is proportional to the radial momentum $P^r$, the radial turning points of the particles can be found by requiring the zeroes of $P^r$ instead of $u^r$. Therefore, the effective potential can be worked out more conveniently by making use of radial momentum $P^r$ \cite{Armaza2016}. By factorizing the energy dependence from the radial dependent part of $(P^r)^2$, one obtains \cite{Jefremov2015,Armaza2016}
\beqn
\fc{(P^r)^2}{m^2}&=&A \bar{e}^2+B\bar{e}+C\nonumber\\
&=&A\left(\bar{e}-\fc{-B+\sqrt{B^2-4AC}}{2A}  \right)\nonumber\\
&&\times\left(\bar{e}-\fc{-B-\sqrt{B^2-4AC}}{2A}  \right),
\label{Motion_Eq}
\eeqn
where for the EiBI black hole \eqref{BH_Metric}, the corresponding parameters $A$, $B$ and $C$ are given by
\beqn
A&=&\frac{4 \left(r^2-M^2 \bar{s}^2 f\right)}{\left[\left(2 r-M^2  \bar{s}^2 f'\right)\psi -2 M^2  \bar{s}^2 f \psi'\right]^2},\\
B&=&\frac{4  \bar{s} \bar{j} M^2  \left[2  \left(\psi -r \psi'\right)f-r \psi f'\right]}{\left[\left(2 r-M^2  \bar{s}^2 f'\right)\psi -2 M^2  \bar{s}^2 f \psi'\right]^2},\\
C&=&\frac{8 M^2 \bar{s}^2 r  f \psi \psi'-4 \left[r^2+M^2 \bar{s}^2 \left(f-r f'\right)\right]\psi^2 }{\left[(2 r-M^2 \bar{s}^2 f')\psi-2 M^2 \bar{s}^2 f \psi '\right]^2}\fc{\bar{j}^2}{\bar{s}^2 }\nn\\
&& +\fc{\bar{j}^2}{\bar{s}^2 }-f.
\eeqn

The effective potential of the spinning test particle is given by the positive square root of \eqref{Motion_Eq}, because the positive square root corresponds to the four-momentum pointing toward future, while the negative corresponds to the four-momentum pointing toward past. So the effective potential reads
\beqn
\!\!\!\!\!\!\!\! V_{\text{eff}} \!&\!\!=\!\!&\! \fc{-B+\sqrt{B^2-4AC}}{2A}\nn\\
\!&\!\!=\!\!&\! \fc{\sqrt{\left(r^2+M^2\bar{j}^2-M^2 \bar{s}^2 f\right)f } }{2 \left(r^2-M^2 \bar{s}^2 f\right)} \big[\left(2 r-M^2 \bar{s}^2 f'\right)\psi \nn\\
&&-2 M^2 \bar{s}^2 f \psi'\big]
+\fc{ M^2\bar{s}\bar{j}  \left[r \psi f'-2 \left(\psi-r \psi '\right)f \right]}{2 \left(r^2-M^2 \bar{s}^2 f\right)}.
\label{Effective_potential}
\eeqn

For simplicity, we only consider the case of asymptotically flat black hole background in rest of the work, i.e., we set $\lambda=1$ so that the cosmological constant $\Lambda$ vanishes. The effective potential for different parameters is illustrated in Fig.~\ref{Fig_Eff_Potential}.

As shown in Fig.~\ref{V_diff_Gt}, the effective potential in EiBI gravity is higher than that in GR for the same spin parameter. Therefore, the same particles with the same initial state move along different trajectories in the two theories, and the distinct would enlarge over time due to increased number of orbital cycles.  By observing the results of Figs.~\ref{V_diff_kappa}, \ref{V_diff_s} and \ref{V_diff_j}, we find that the effective potential decreases with the spin angular momentum $\bar s$ but increases with the deviation parameter $\kappa$ and the total angular momentum $\bar j$ (or equivalently orbital angular momentum $\bar l$ defined by $\bar l=\bar j-\bar s$). Here we notice again that the potentials will reduce to that of RN black hole in GR in the limit $\kappa\to 0$ from Fig.~\ref{V_diff_kappa} .

\begin{figure}[t]
\begin{center}
\subfigure[$|\dot r|$]  {\label{Radial_Velocity}
\includegraphics[width=4.1cm,height=3.5cm]{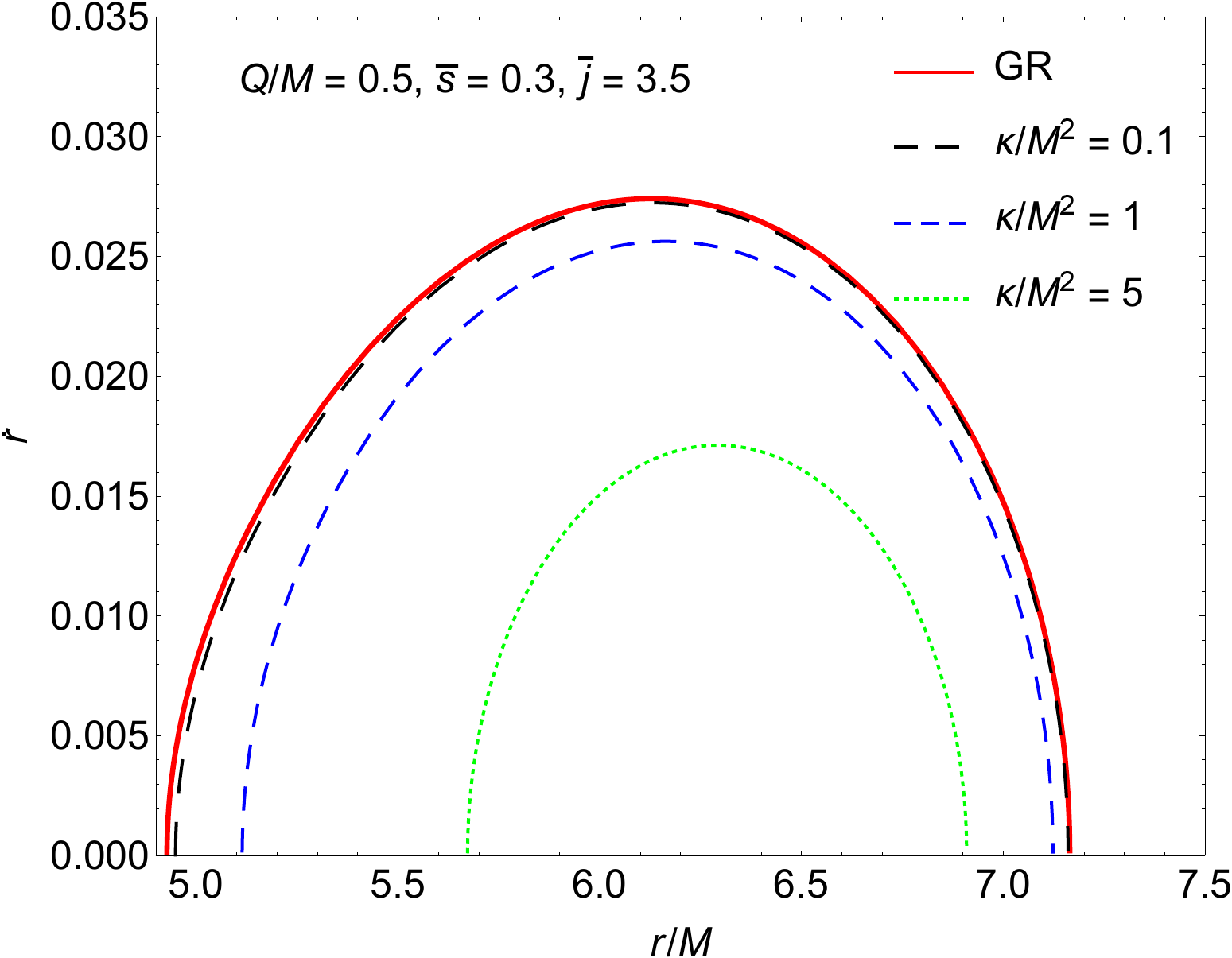}}
\subfigure[$\dot \phi$]  {\label{Angular_Velocity}
\includegraphics[width=4.1cm,height=3.5cm]{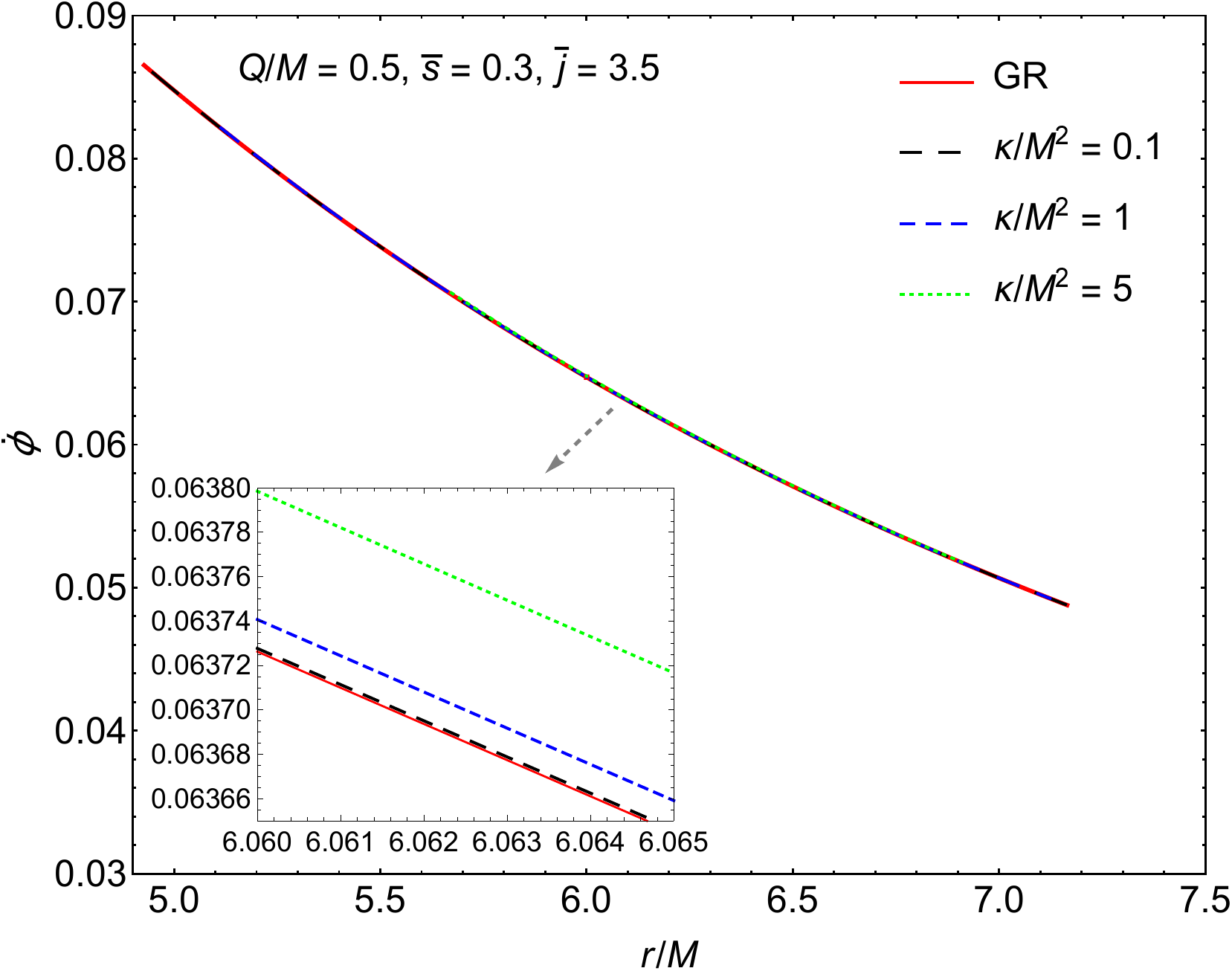}}
\end{center}
\caption{The shapes of the radial velocity $\dot r$ and angular velocity $\dot \phi$ for different $\kappa$.}
\label{Fig_Orbit_Velocities}
\end{figure}

\begin{figure}[t]
	\begin{center}
			\includegraphics[width=4.1cm,height=3.5cm]{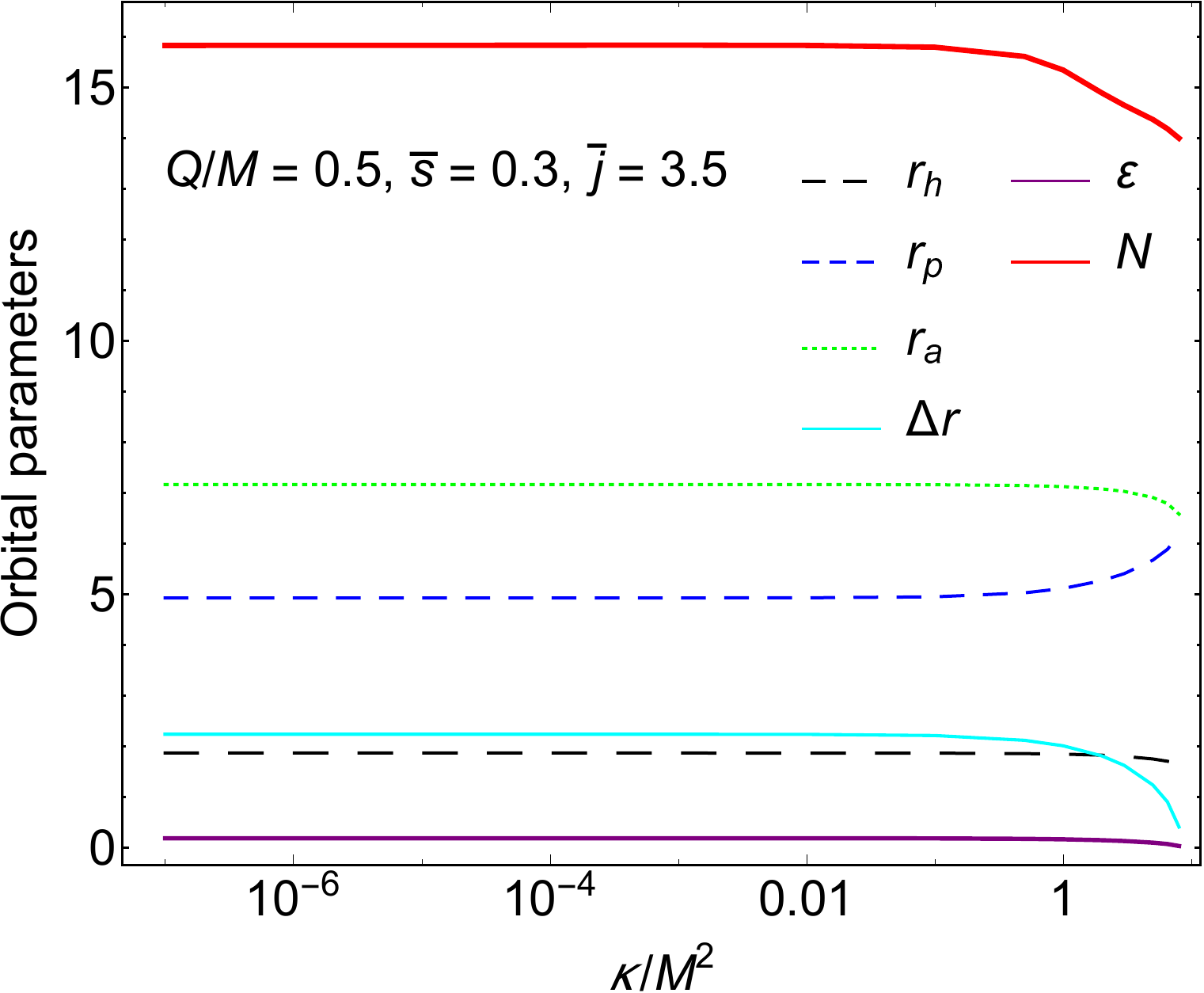}
	\end{center}
	\caption{The dependency between the orbital parameters and the deviation parameter $\kappa$.}
	\label{Orbital_par}
\end{figure}

In particular, as shown in Fig.~\ref{Fig_Orbit}, for a classical test particle trapped in the potential well with energy $\bar e=0.9365$, the particle has lower radial speed and narrower radial range of motion as $\kappa$ increases, which can be seen clearly through Fig.~\ref{Radial_Velocity} as well. So the orbit varies with the deviation parameter $\kappa$. The distinct can be revealed more clearly if we release all particles at their apoapsis of orbits and trace their trajectories in a same amount of time. By integrating the formulas of radial velocity \eqref{ur} and angular velocity \eqref{uphi} numerically to get the radial displacement and angular displacement, we plot the trajectories of these particles in Figs.~\ref{Fig_Orbit} (e-h). {Especially, the corresponding orbital parameters are listed in Tab.~\ref{Tab_Orbits}. Moreover, in order to show more clearly the dependency between the orbital parameters and the deviation parameter $\kappa$,  we plot their relations in Fig. \ref{Orbital_par}}.

 \begin{table*}[t]
\begin{center}
    \begin{tabular}{|C{2cm}|C{2cm}|C{2cm}|C{2cm}|C{2cm}|C{2cm}|C{2cm}|}
     \hline
     $\kappa/M^2$ &  $r_h$ &  $r_a/M$ &  $r_p/M$ &  $\Delta r/M$   & $\varepsilon$  & $N$    \\
     \hline
      GR  & 1.8660  & 7.1649 & 4.9275  & 2.2374 & 0.1850  & 15.8275    \\
       \hline
      0.1  & 1.8640  & 7.1609 & 4.9487  &2.2122 & 0.1827  & 15.8002    \\
       \hline
       1  & 1.8456  & 7.1234 & 5.1139 &  2.0095 & 0.1642  & 15.3260    \\
       \hline
       5  & 1.7495  & 6.9095 & 5.6724  & 1.2371 & 0.0983  & 14.3690    \\
       \hline
\end{tabular}\\
\caption{The horizon radius $r_h$, apastron $r_a$, periastron $r_p$, radial range of motion $\Delta r$, eccentricity $\varepsilon$, and number of rotations $N$ associate to the orbits in Fig.~\ref{Fig_Orbit}.}
\label{Tab_Orbits}
\end{center}
 \end{table*}
As a result, the apastron decreases with $\kappa$  and periastron increases with $\kappa$, therefore, the radial range of motion $\Delta r$ reduces with $\kappa$. The shape of orbit can be characterized by making use of the orbital eccentricity $\varepsilon$, defined as $\varepsilon=\fc{r_a-r_p}{r_a + r_p}$, with $r_a$ and $r_p$ the apastron and periastron of the orbit respectively. Table~\ref{Tab_Orbits} and Fig. \ref{Orbital_par} revel that the orbital eccentricity reduces with the deviation parameter $\kappa$.

Interestingly, as  $\kappa$ increases, the particle rotates less number of circles in the same amount of time. This is due to the fact that the particle with larger $\kappa$ rotates, on average, at more distant orbits, and the angular velocities decrease more sensitively with the radius $r$ but increase less sensitively with $\kappa$, see Fig.~\ref{Angular_Velocity}.

{It is noted that all the orbits illustrated in Fig.~\ref{Fig_Orbit} and Tab.~\ref{Tab_Orbits} are chosen in the vicinity of the black hole. To obtain the constraint for the parameter $\kappa$, one can use the data of the orbits for the observed stars around the Sagittarius A*. In astronomy, the observed stars are far from the central black holes, such as the periastron $r_p/M$ and  apastron $r_a/M$ of observed star S2 around the Sagittarius A* are $(893.285M,  24145.9M)$ \cite{Gillessen:2008qv}. Note that, the magnitude of the spin for the test particle is dependent on the mass ratio $m/M$ and it is about $10^{-6}$ \cite{Zhang2018}. The corresponding orbital angular momentum for such orbit is about $\bar{j}\sim 50$. Therefore, the true value of the spin for the test particle is really tiny and spin-curvature force can be ignored for the observed orbits \cite{Gillessen:2008qv} around Sagittarius A*. Besides the orbital and spin angular momentum of the test particle, the spin angular momentum of the central supermassive black hole can also affect the orbits of the stars around itself. However, there are no solid measurements for the spin of the supermassive black hole in the center of Sagittarius A* and the spin can be in the range of (0.1,0.98) \cite{Rockefeller:2005ta,Prescher:2005sd,Broderick:2008sp,Broderick:2010kx,Sanjeev2021}. Under such larger error of spin, we only consider the spinless case for the central spuermassive black hole to naively estimate the magnitude of $\kappa$.
		
For the orbits with large radii, we work perturbatively and obtain the analytic formula to estimate the deviation induced by the deviation parameter $\kappa$. The leading influence of deviation parameter $\kappa$ on effective potential $V_{\text{eff}}$ is given by $\delta V_{\text{eff}} \approx\frac{\kappa Q^2}{2 r^4}$. Correspondingly, the changes of the apastron $r_a$ and periastron $r_p$ are approximated analytically as
\beq
\frac{\delta r_{a,p}}{r_{a,p}} \approx -\fc{\kappa Q^2}{2 M \lt({r_{a,p}}-M \bar{j}^2\rt) r_{a,p}^2}.
\label{delta_radius}
\eeq
In principle, the magnitude of above deviation \eqref{delta_radius} of the orbit induced by the non-zero $\kappa$ should be smaller than the observation error of the observed orbits around Sagittarius A* \cite{Gillessen:2008qv}. To obtain the constraint for the parameter $\kappa$, we naively let the error of the orbits as the maximum of the orbital deviations \eqref{delta_radius}. Checking the observed orbits for the stars around the Sagittarius A*, one can find that the observation data for the orbit of the S2 star is the most accurate \cite{Gillessen:2008qv}. Therefore, we compute the constraint on parameter $\kappa$ in terms of the observed orbit of the S2 star around the Sagittarius A*.

For the S2 star, the semi major axes of the observation is $a_{\text{orb}}=0.123''\pm0.001''$ and the eccentricity is $\varepsilon_{\text{orb}}=0.880\pm0.003$, where $'$ is the arc angle. For which, the corresponding length for a arc angle is about $1.2\times 10^{15}m$. With the help of the orbital parameters of S2 star, we can get the corresponding periastron $r_p$ and apastron $r_a$ as follows
\beq
r_{a,p}=a_{\text{orb}}(1\pm \varepsilon_{\text{orb}}).
\label{orbit_radius_s2}
\eeq
Based on the Eq. \eqref{orbit_radius_s2}, the error of $r_{a,p}$ can be obtained as follows
\beq
\frac{\delta r_{a,p}}{r_{a,p}}=\frac{\delta a_{\text{orb}}}{a_{\text{orb}}}\pm \frac{\delta \varepsilon_{\text{orb}}}{1\pm \varepsilon_{\text{orb}}}.
\eeq
We have shown that the non-zero $\kappa$ increases the periastron $r_p$ and decreases the apastron $r_a$. Substituting the values of the error for the orbital parameters of S2 star, we have
\beqn
\frac{\delta r_{a}}{r_{a}}&\simeq& \frac{0.001}{0.123} + \frac{0.003}{1+0.880}\simeq0.0097,\\
\frac{\delta r_{p}}{r_{p}}&\simeq& \frac{0.001}{0.123} - \frac{0.003}{1-0.880}\simeq-0.01687.
\eeqn
References \cite{Zajacek:2018vsj,Zajacek:2018ycb} have shown that the upper observational limit on the charge of galactic central black hole is on the order of $Q \leq 3\times 10^8C$, the corresponding value in the natural unit is about $Q/M\sim10^{-11}$ for a supermassive black hole with a mass $10^6 M_\odot$, where $M_\odot$ is the solar mass. Combining the magnitudes of $\frac{\delta r_{a}}{r_{a}}$, $\frac{\delta r_{p}}{r_{p}}$, and the charge of galactic central black hole, we find that the most stringent constraint on the upper limit of $\kappa$ is in the magnitude of $10^{29}$ at least. Under such larger magnitude for parameter $\kappa$,  we conclude that we can not obtain the serious constraint for the parameter $\kappa$ in terms of the observed orbital parameters around Sagittarius A* \cite{Gillessen:2008qv}. To obtain more accurate constraint, the orbits in the vicinity of the black hole are necessary, and it is the reason that we only consider the orbits with small radii.
}

\subsection{Causality}

\begin{figure}[t]
\begin{center}
\subfigure[EiBI balck hole]  {\label{Causality_EiBI}
\includegraphics[width=4.1cm,height=3.5cm]{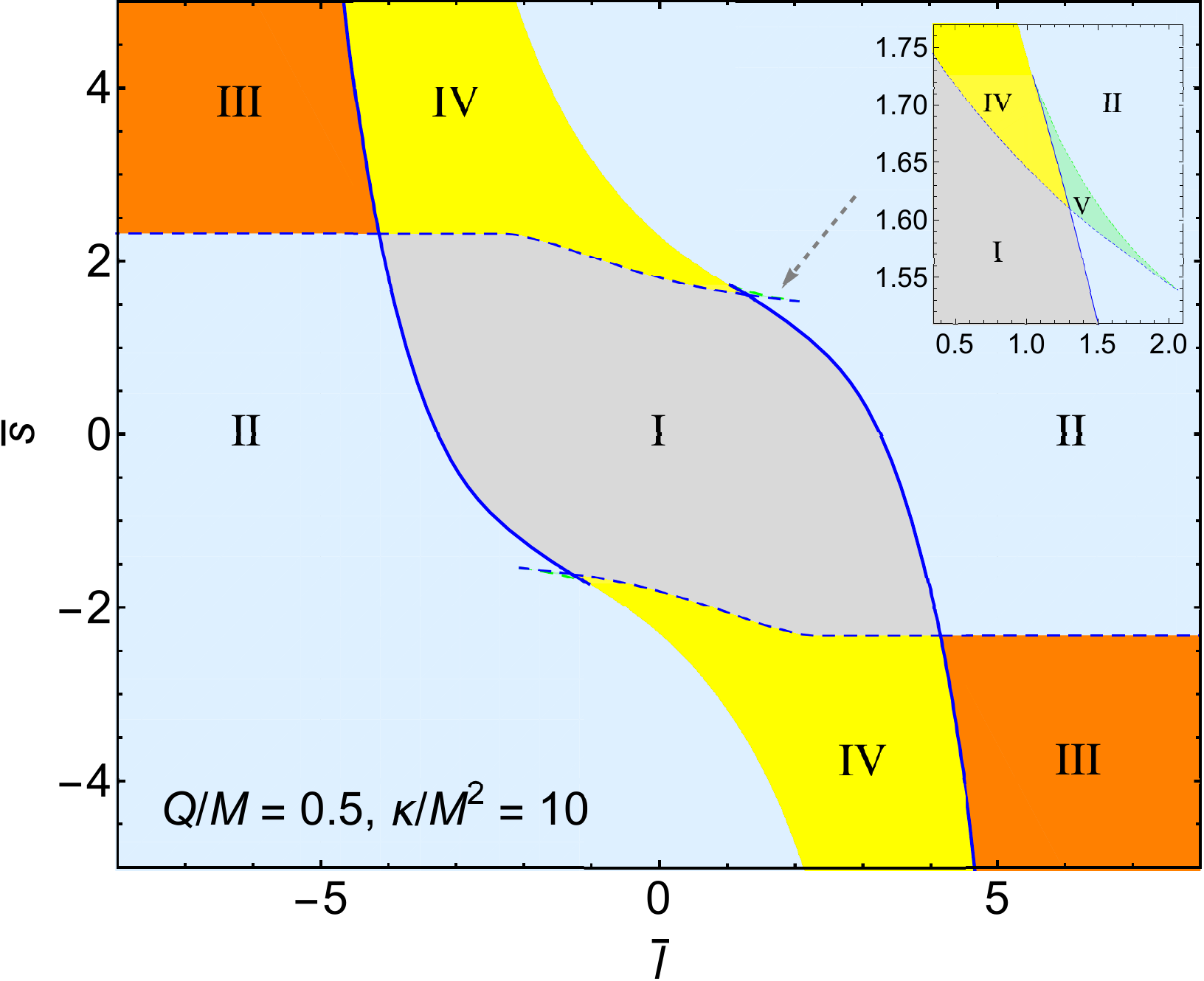}}
\subfigure[RN black hole]  {\label{Causality_GR}
\includegraphics[width=4.1cm,height=3.5cm]{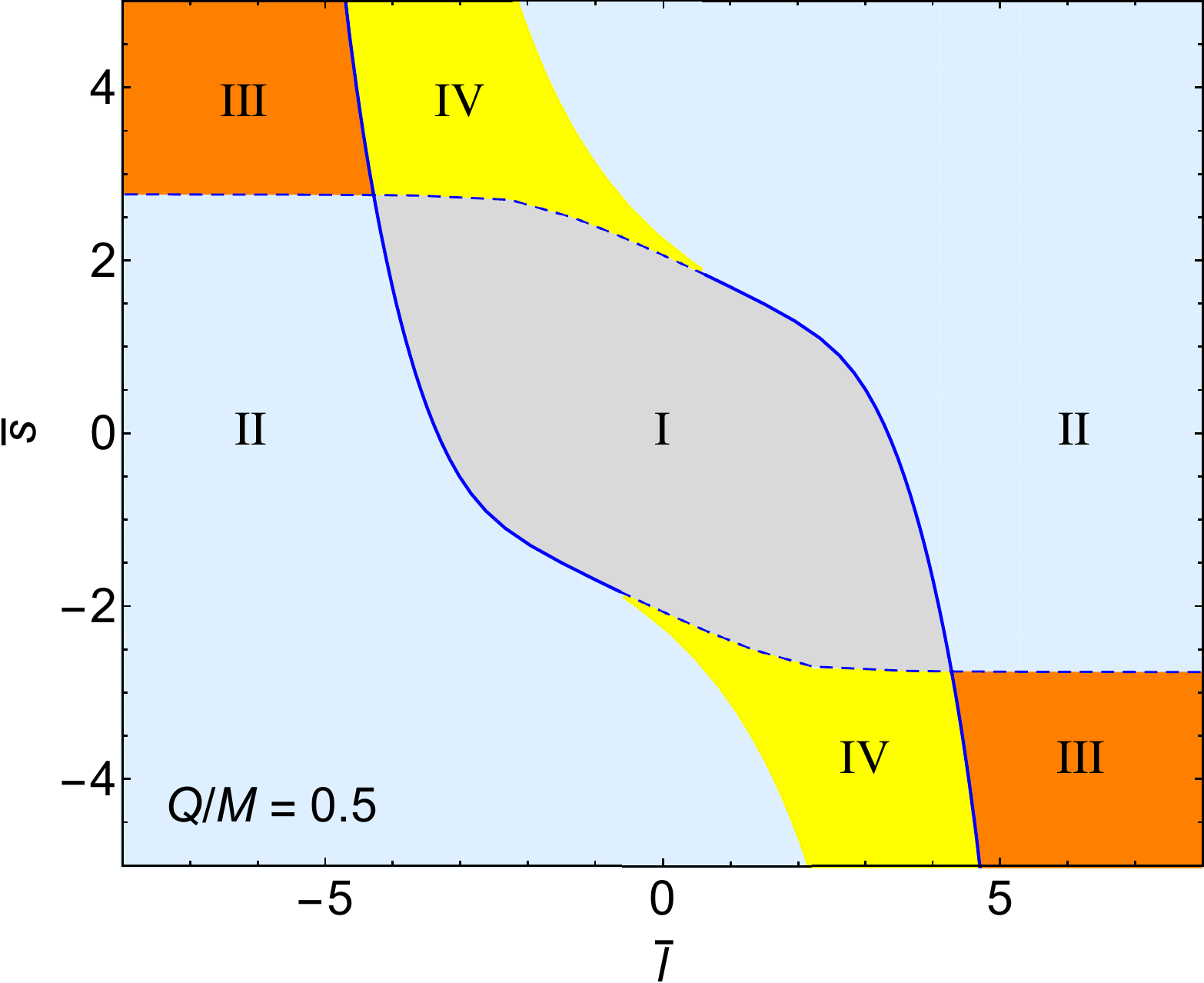}}
\end{center}
\caption{The properties of the circular orbits for the spinning test particle in EiBI black hole and RN black hole. The gray colored region I refers to the parameters without stable circular orbit. The light blue colored regions II refers to the parameters with only one stable timelike circular orbits. The yellow colored regions IV refer to the parameters with only spacelike and unphysical circular orbits. The orange colored region III and green colored region V refer to the parameters with two stable circular orbits.}
\label{Causality}
\end{figure}

Since the four-velocity $u^\mu$ is generally not parallel to the four-momentum $P^\mu$, the trajectory may transform from timelike to spacelike. So it is necessary to check the causality of circular orbits by using the superluminal constraint condition \eqref{Superluminal_Constraint}. We numerically scan the causality in the parameter space $(\bar l, \bar s)$ for EiBI black hole with $\kappa/M^2=10$. And for comparison, we scan the causality for RN black hole as well. The final results are illustrated in Fig.~\ref{Causality}. For the causality of particle's circular orbits in  EiBI black hole, its properties are summarized as follows:
\begin{itemize}
	{\item In region I, the particle does not have any stable circular orbit.}
	{\item In regions II, the particle has only one stable circular orbit, and the orbit is subluminal and physical.}
	{\item In regions IV the particle has only one stable circular orbit, however, the orbit is superluminal and unphysical.}
	{\item In regions III and V, the particle has two stable circular orbits, however, the inner orbit is superluminal and unphysical, and only the outer orbit is subluminal and physical.}
\end{itemize}

It is observed that at the junctions of regions I, II and IV in  Fig.~\ref{Causality_EiBI}, two new green colored regions marked as V emerge in EiBI black hole, where the particle has two stable circular orbits with one subluminal and the other superluminal. The orbital property in regions V is similar to that in regions III, because they are both the cross regions between two lines of ISCO.

\subsection{ISCO}

Among the trajectories of particles, there is a special kind of orbits which refers to stable circular orbits. They satisfy the conditions:  (i) $\fc{dr}{d\lambda}=0$, i.e., its radial velocity vanishes, and (ii) $\fc{d^2r}{d\lambda^2}=0$ (or $\fc{dV_{\text{eff}}}{dr}=0$), i.e., its radial acceleration (or effective force) vanishes. In order to keep the orbit stable to against small fluctuations, another condition,  (iii)  $\fc{d^2V_{\text{eff}}}{dr^2}>0$, is required. So the particle just rotates along the local minimum of the effective potential well. The minimum radius of these stable circular orbits is nothing but the so-called ISCO.  It depends on the mass, charge, spin and deviation parameters of the background black hole, and especially, it marks the inner edge of black hole accretion disk, so it is important from the observational point of view.

\begin{figure}[t]
	\begin{center}
		\subfigure[$\kappa/M^2=1$]  {\label{ISCO_parameter_kappa1}
			\includegraphics[width=4.1cm,height=3.5cm]{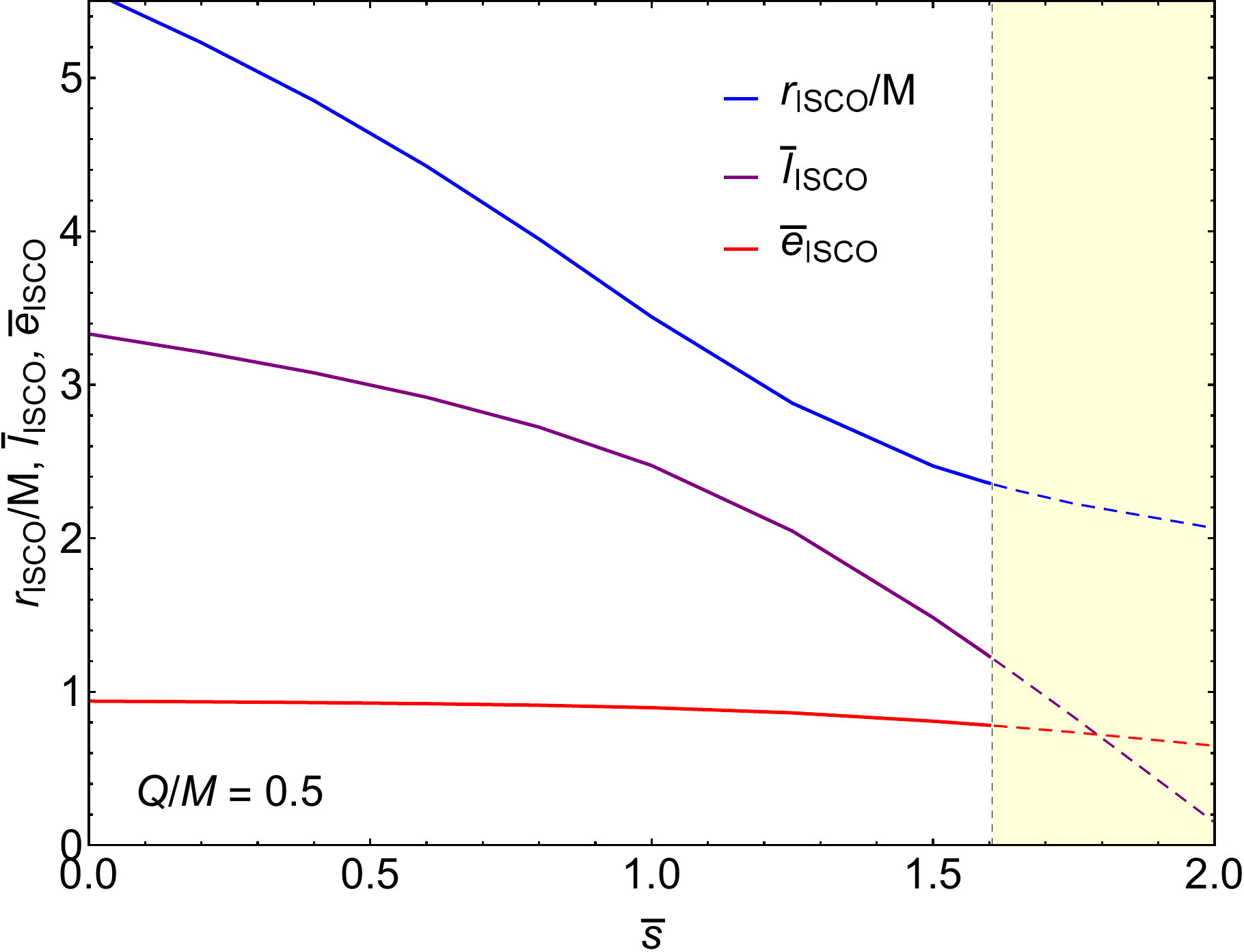}}
		\subfigure[$\kappa/M^2=5$]  {\label{ISCO_parameter_kappa5}
			\includegraphics[width=4.1cm,height=3.5cm]{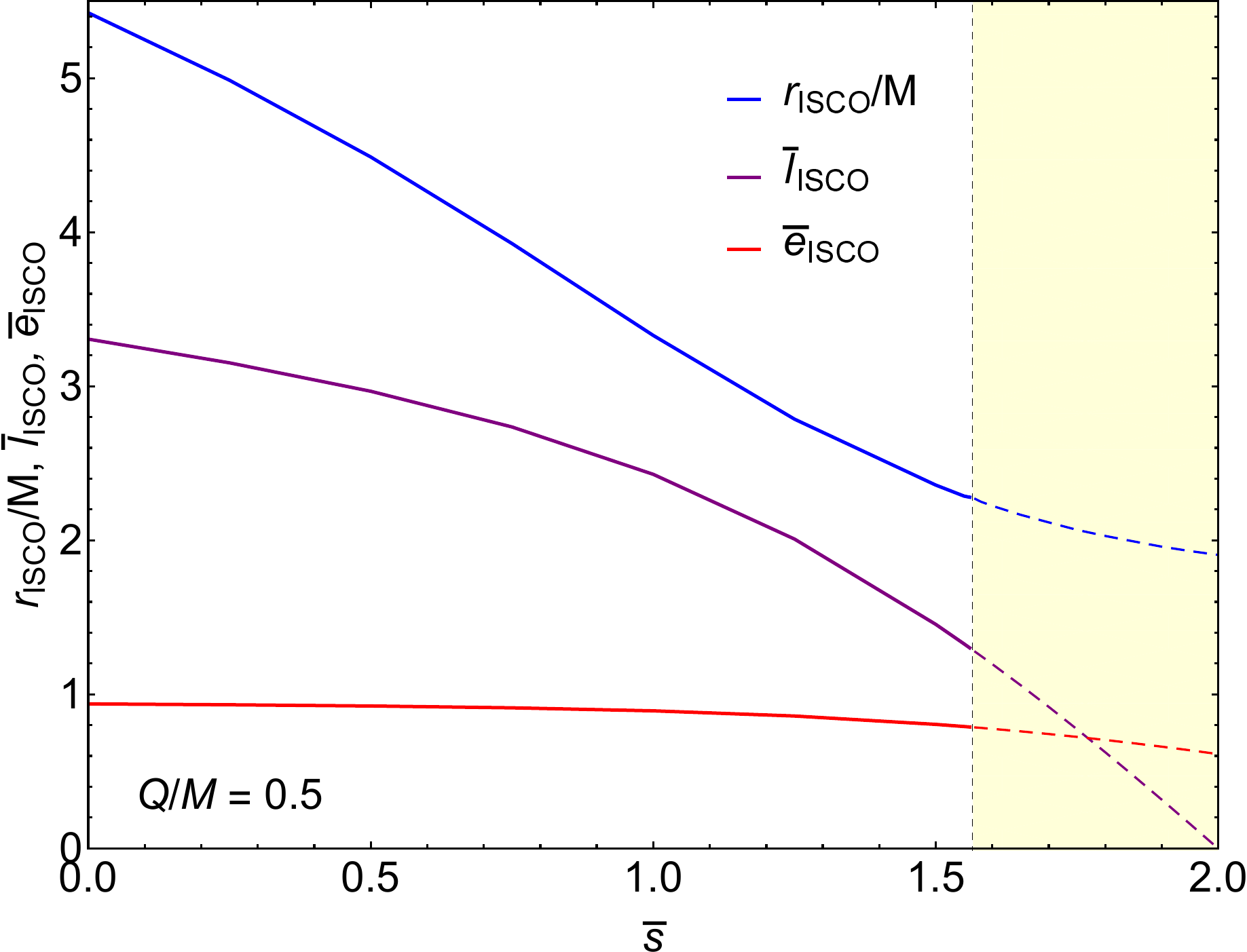}}\\
		\subfigure[$\kappa/M^2=10$]  {\label{ISCO_parameter_kappa10}
			\includegraphics[width=4.1cm,height=3.5cm]{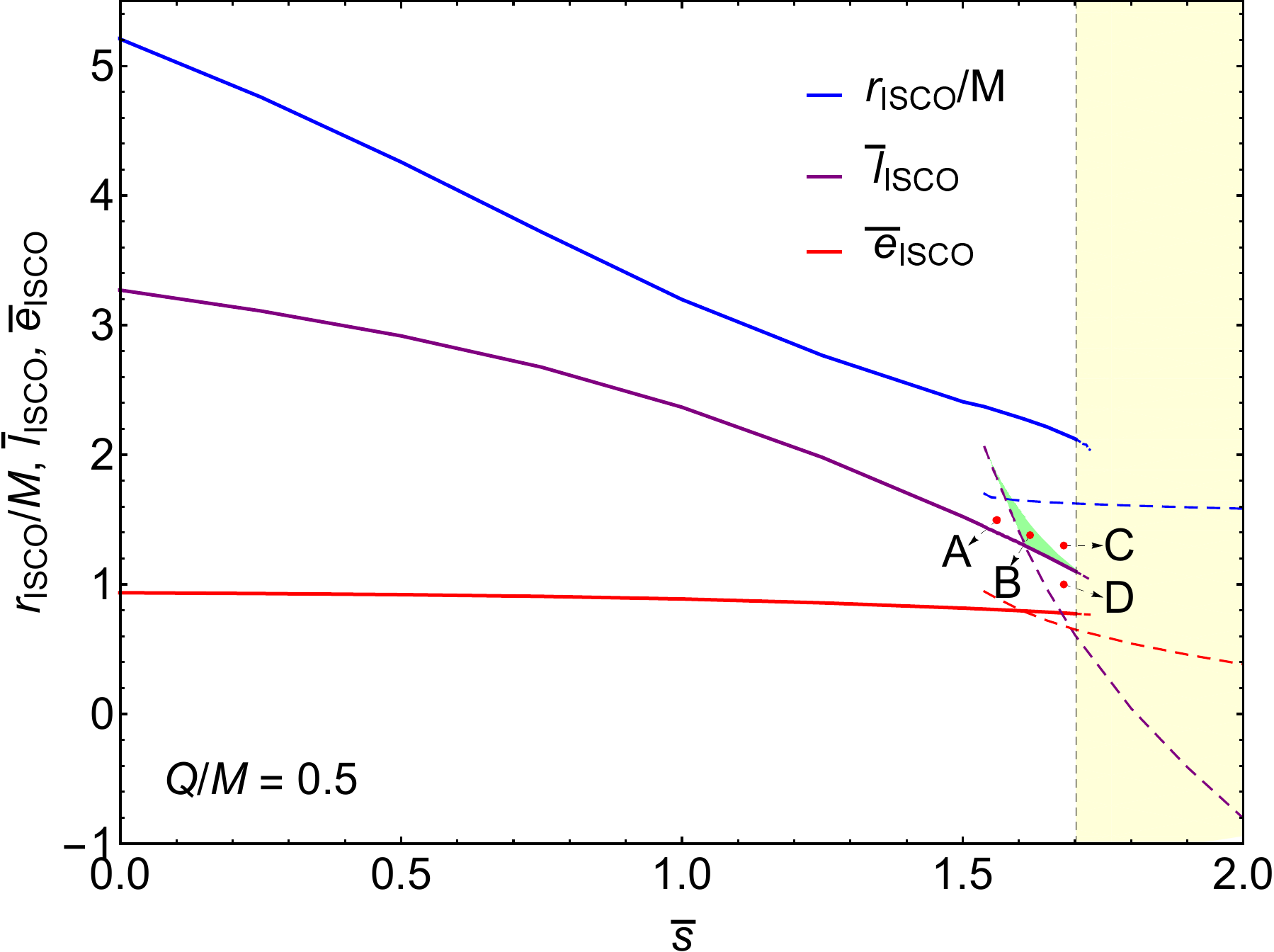}}
		\subfigure[$\bar{s}=0$]  {\label{ISCO_parameter_s0}
			\includegraphics[width=4.1cm,height=3.5cm]{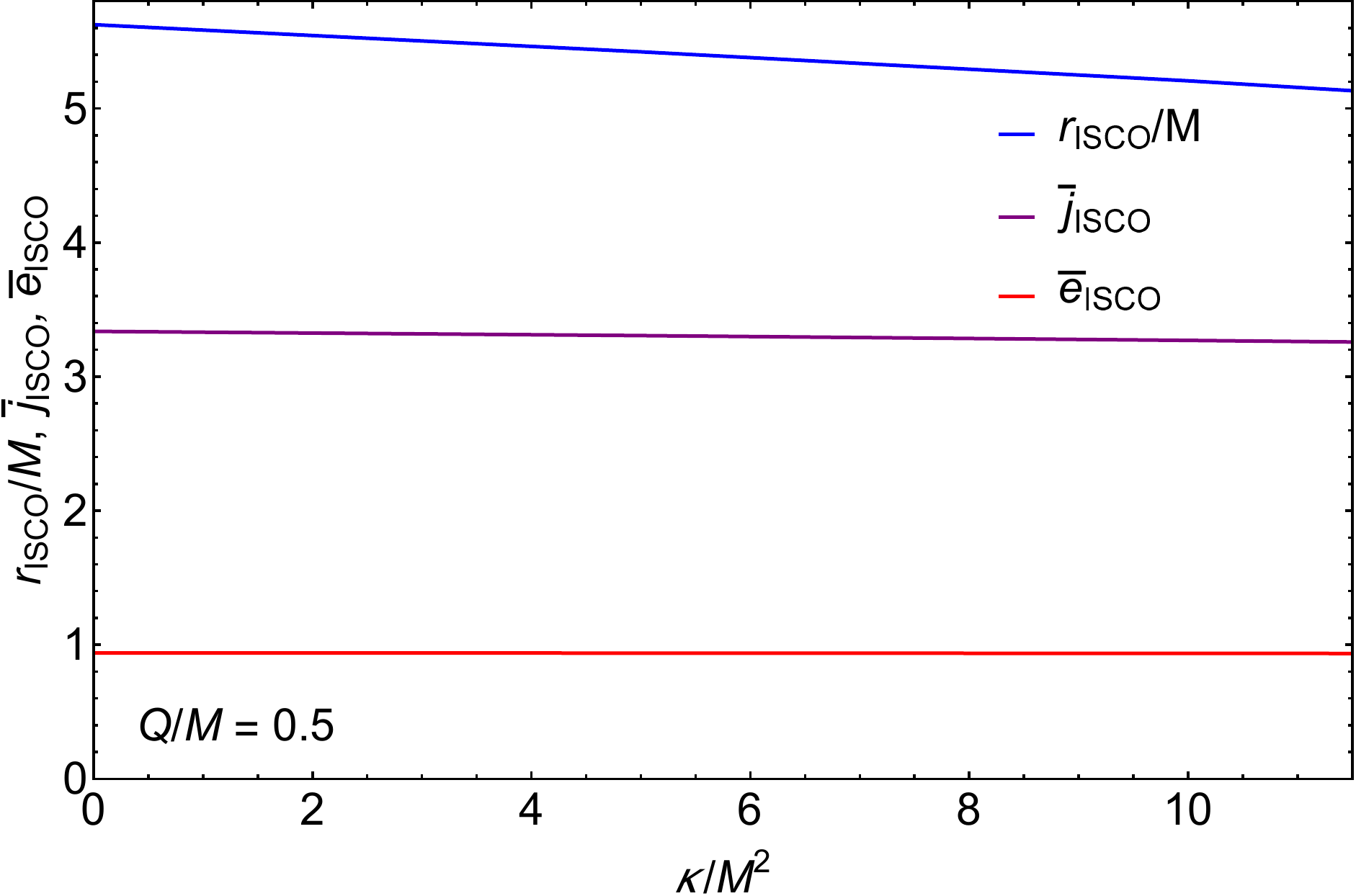}}\\
		\subfigure[$\bar{s}=1$]  {\label{ISCO_parameter_s1}
			\includegraphics[width=4.1cm,height=3.5cm]{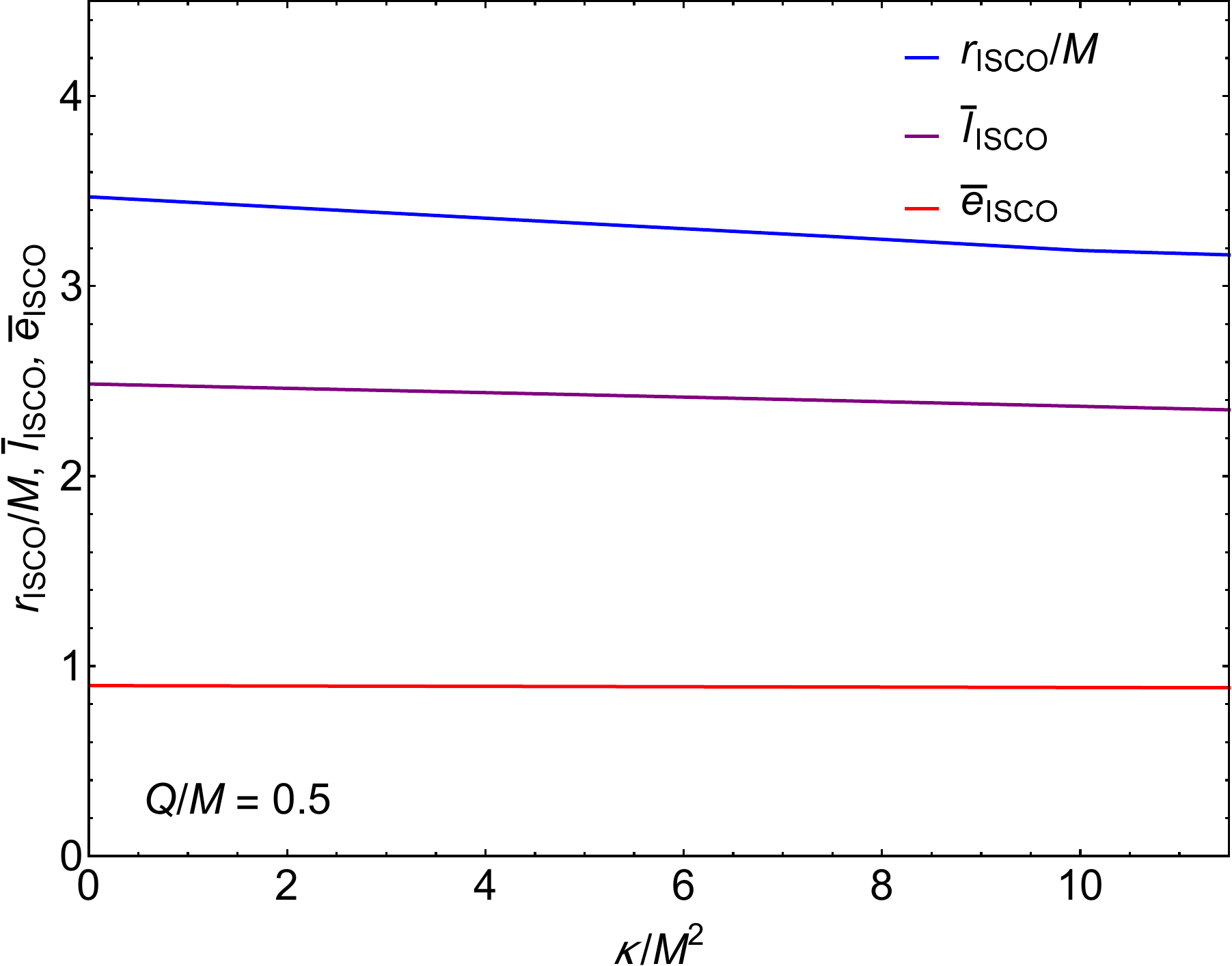}}
		\subfigure[$\bar{s}=1.6$]  {\label{ISCO_parameter_s16}
			\includegraphics[width=4.1cm,height=3.5cm]{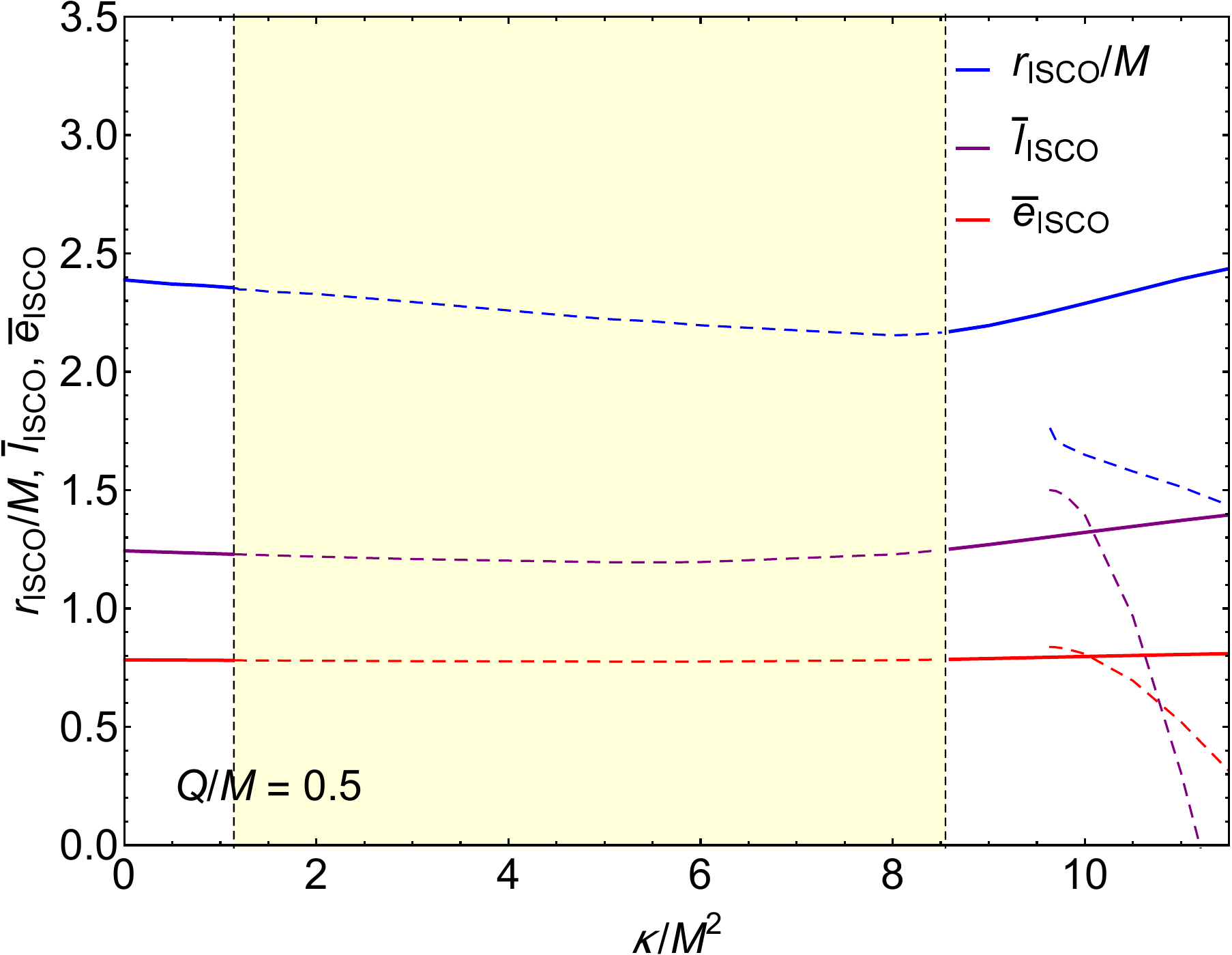}}
	\end{center}
	\caption{The radius $r_\text{ISCO}$, orbital angular momentum $\bar{l}_\text{ISCO}$ and energy $\bar{e}_\text{ISCO}$ of the ISCO for different parameters $\kappa$ and $\bar{s}$, where the thick lines refers to the timelike ISCO and dashed to spacelike. The areas colored in light yellow represents the region without timelike ISCO.}
	\label{ISCO_parameter}
\end{figure}

\begin{figure}[htb]
	\begin{center}
		\subfigure[Point A]  {\label{ISCO_parameter_Potential_A}
			\includegraphics[width=4.1cm,height=3.5cm]{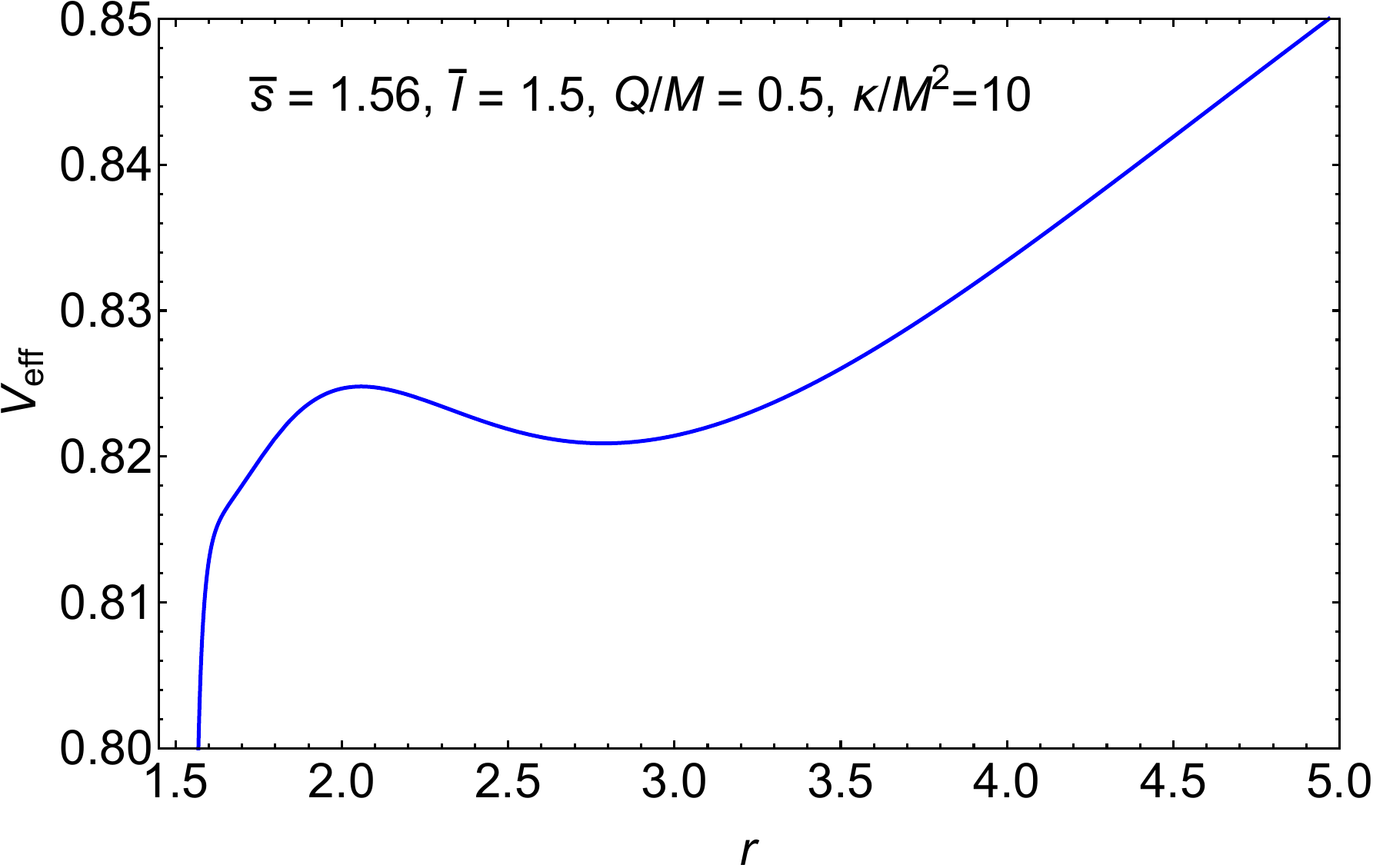}}
		\subfigure[Point B]  {\label{ISCO_parameter_Potential_B}
			\includegraphics[width=4.1cm,height=3.5cm]{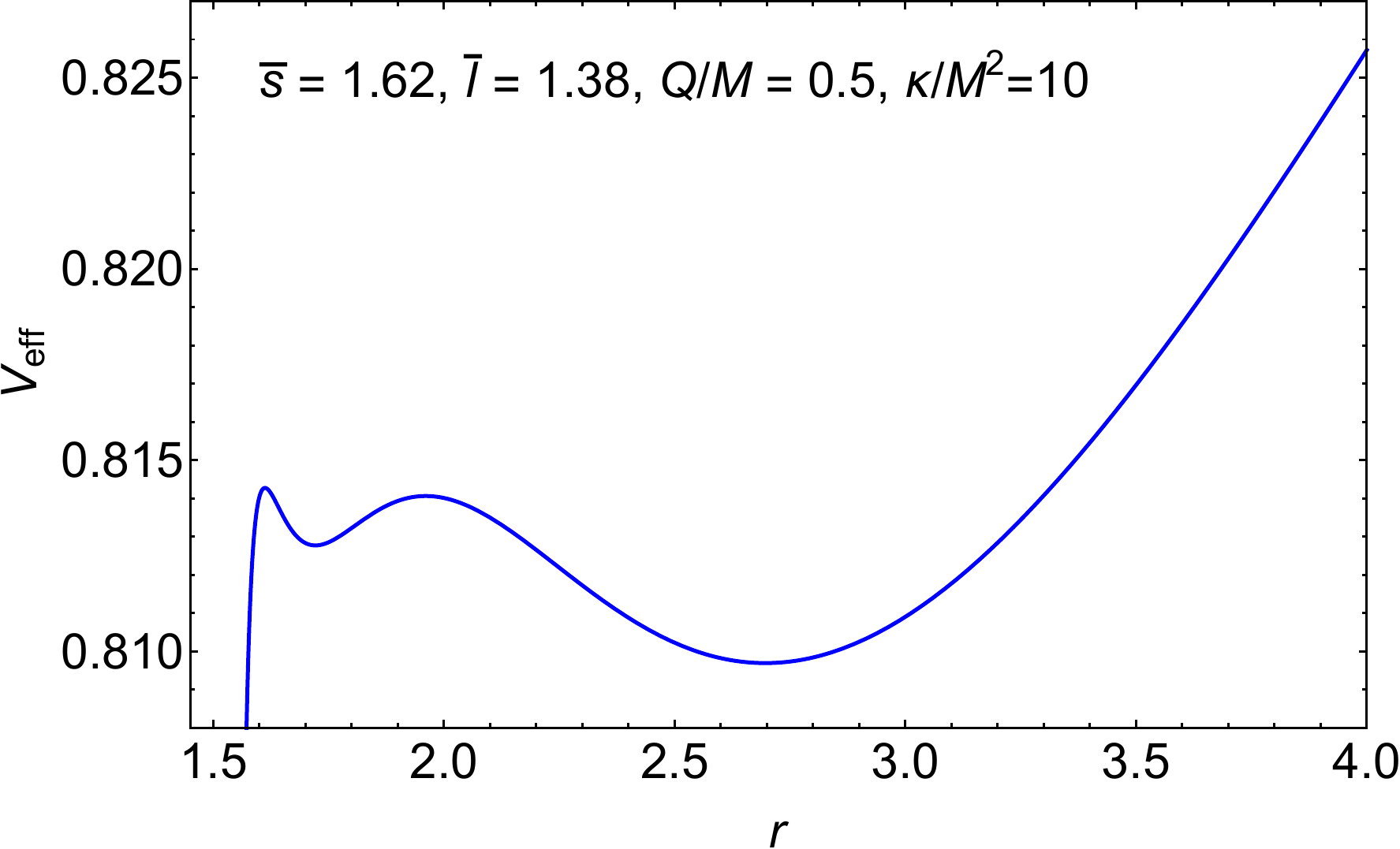}}\\
		\subfigure[Point C]  {\label{ISCO_parameter_Potential_C}
			\includegraphics[width=4.1cm,height=3.5cm]{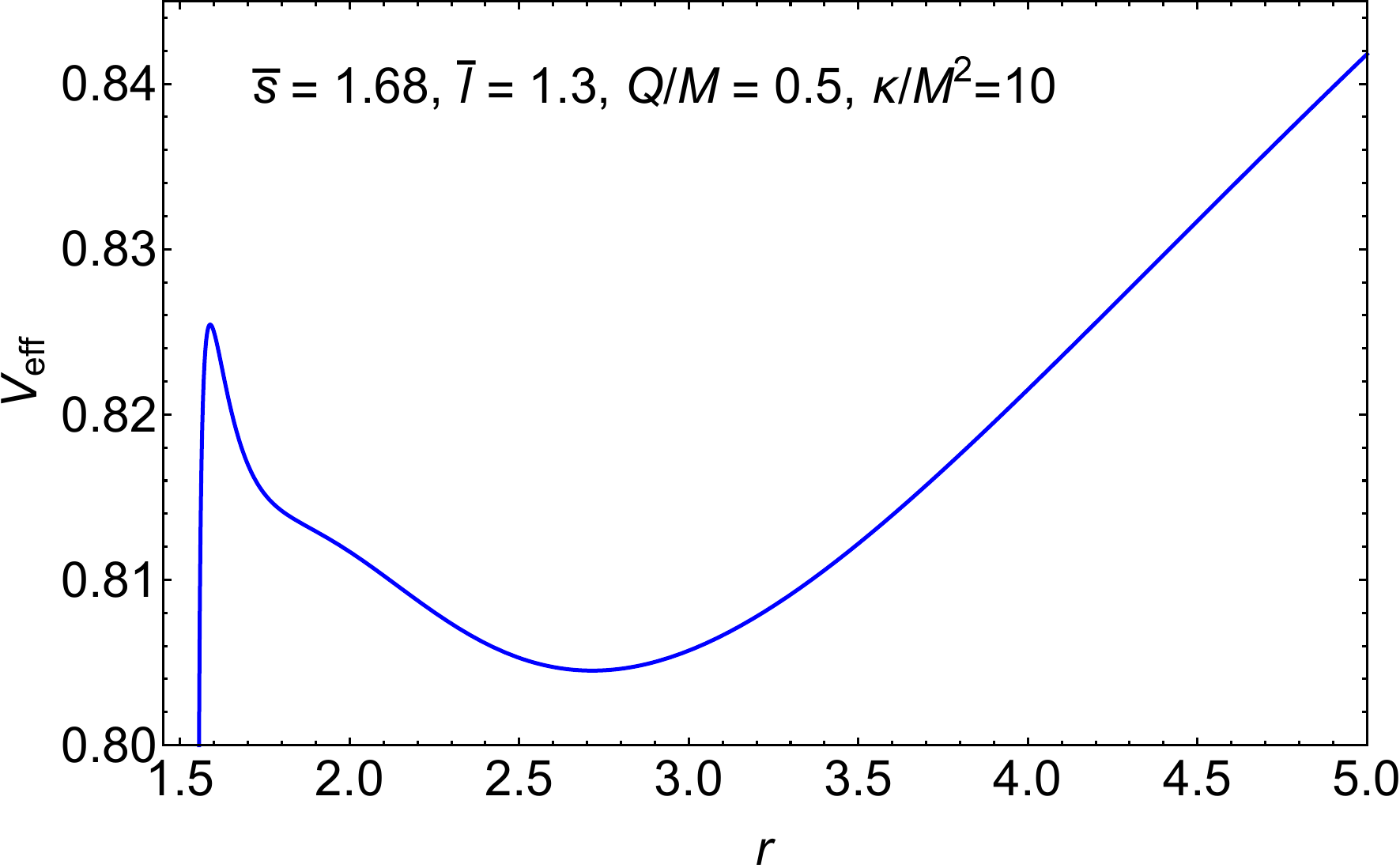}}
		\subfigure[Point D]  {\label{ISCO_parameter_Potential_D}
			\includegraphics[width=4.1cm,height=3.5cm]{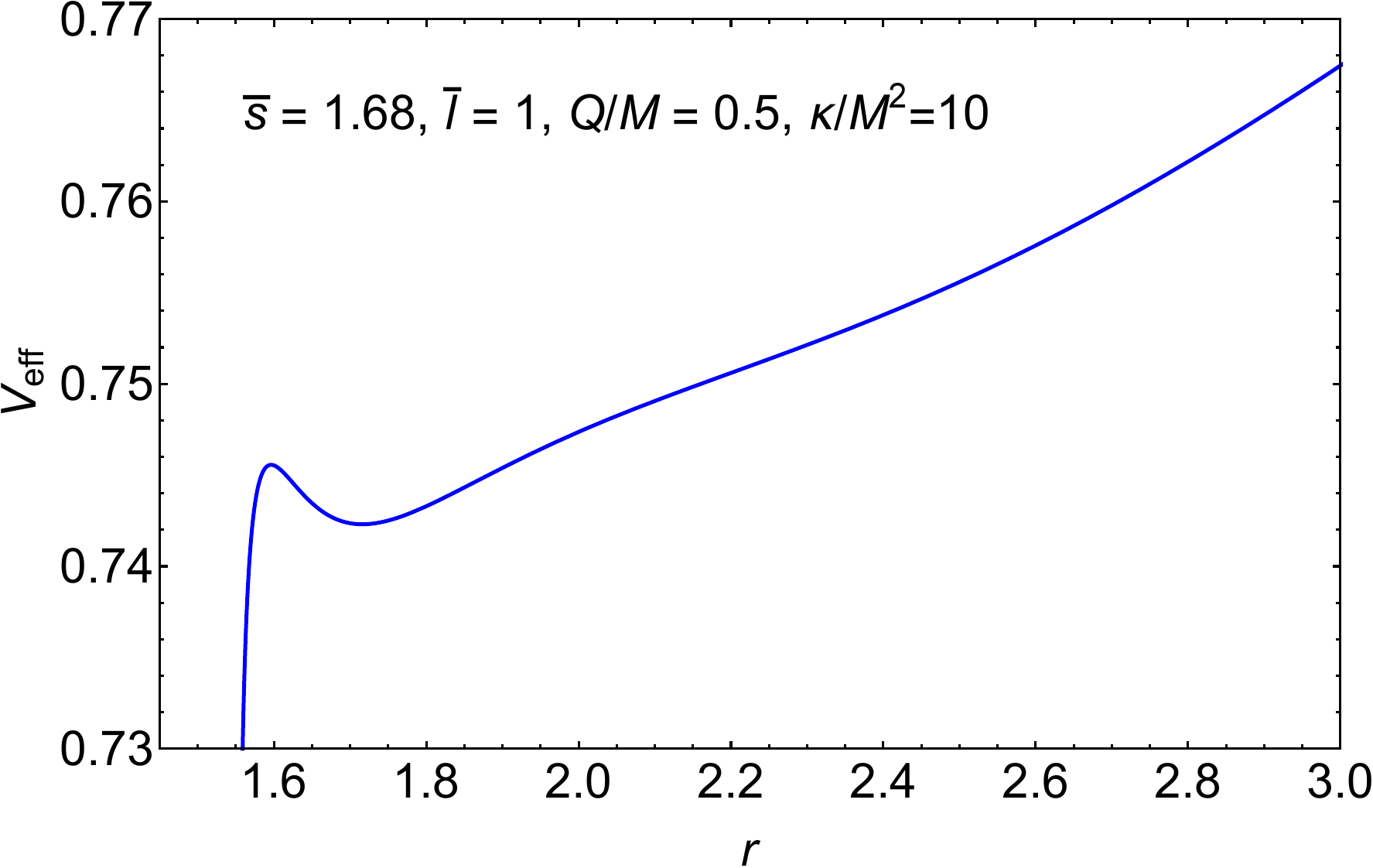}}
	\end{center}
	\caption{The shapes of potential referring to points A, B, C and D in  Fig.~\ref{ISCO_parameter_kappa10}.}
	\label{ISCO_Potential_Points}
\end{figure}

In Fig.~\eqref{ISCO_parameter}, we plot the radius $r_\text{ISCO}$, orbital angular momentum $\bar{l}_\text{ISCO}$ and energy $\bar{e}_\text{ISCO}$ of ISCO for different deviation parameter $\kappa$ and spin $\bar{s}$. The causality property of ISCO is encoded in it as well. From Fig.~\ref{ISCO_parameter_kappa10}, it is easy to observe that the borderlines of regions I, III and V in Fig.~\ref{Causality} are just the ISCO, where the thick lines refers to the subluminal ISCO and dashed to superluminal. The numerical results tell us that the ISCO is subluminal and physical when the spin $\bar s$ is small, however it may become superluminal when the spin increases. Furthermore, the causality property of particles in EiBI black hole is similar to that in RN black hole when the deviation parameter $\kappa$ is small, and it will recover exactly the result of RN black hole as $\kappa\to 0$. However, the causality property becomes more complex when $\kappa$ goes large enough, just as shown in Figs.~\ref{Causality_EiBI}, \ref{ISCO_parameter_kappa10} and \ref{ISCO_parameter_s16}.

For the spin $\bar s$ dependence of ISCO, the result of Figs.~\ref{ISCO_parameter} (a-c) revels that all the radius $r_\text{ISCO}$, $\bar{l}_\text{ISCO}$ and $\bar{e}_\text{ISCO}$ decrease with $\bar s$. This is a well known result for spinning test particles moving in black holes of GR \cite{Jefremov2015,Zhang2017}. For small $\kappa$, the spin dependence shown in Figs.~\ref{ISCO_parameter_kappa1} and \ref{ISCO_parameter_kappa5} is quit similar to that of RN black hole in GR \cite{Zhang2017}. However, by observing Fig.~\ref{ISCO_parameter_kappa10}, an obvious difference emerges for large deviation parameter $\kappa$, i.e., there are two possible ISCOs for $1.539<\bar s<1.726$ with $\kappa/M^2=10$. The typical shapes of effective potentials around the green colored region in Fig.~\ref{ISCO_parameter_kappa10} are illustrated in Fig.~\ref{ISCO_Potential_Points}. As shows in Fig.~\ref{ISCO_parameter_Potential_B}, the effective potential has two local minima inside the green colored region, therefore, it has two stable circular orbits. The two possible ISCOs emerge because the one (thick lines) involves to the critical points that the outer potential well becomes flat, while the other (dash lines) involves to the critical points that the inner potential well becomes flat. However, because the inner one of the two stable circular orbits is superluminal, only the outer ISCO is physical.

For the deviation parameter $\kappa$ dependence of ISCO, Figs.~\ref{ISCO_parameter_s0} and \ref{ISCO_parameter_s1} show that $r_\text{ISCO}$, $\bar{l}_\text{ISCO}$ and $\bar{e}_\text{ISCO}$ decrease with $\kappa$ when the particle's spin is small. However, as shown in Fig.~\ref{ISCO_parameter_s16}, when the spin becomes large enough,  $r_\text{ISCO}$,  $\bar{l}_\text{ISCO}$ and $\bar{e}_\text{ISCO}$ will first decrease and then increase with $\kappa$. Especially, for $\bar s=1.6$, there is  another possible ISCO when $\kappa/M^2>9.641$, which is superluminal and decreases with $\kappa$. The constraint from the existence of neutron stars suggests $|\kappa|\lesssim 1 \text{m}^5 \text{kg}^{-1} \text{s}^{-2}$ \cite{Pani2011}, i.e., $\lt | {\kappa}/{M^2} \rt |\lesssim 6.87 \times 10^3 \times \lt(\fc{M_\odot}{M} \rt)^2$ \cite{Sotani2014}. For the mass $M$ of intermediate-mass black holes ($10^2 M_\odot-10^5 M_\odot$) or supermassive black holes ($10^5 M_\odot-10^9 M_\odot$), $\lt | {\kappa}/{M^2} \rt |$ must be small, thus the ISCO radius $r_\text{ISCO}$ decreases monotonously with $\kappa$, and it leads to a smaller inner edge of accretion disk around the EiBI black hole than that around the RN black hole.  For the mass $M$ of stellar-mass black holes ranging from about 5 to several tens of solar masses, the upper limit of $\lt | {\kappa}/{M^2} \rt |$ is ranging from about 1 to several hundreds. In this case, the inner edge of accretion disk around the EiBI black hole may greater than that around the RN black hole if $\kappa/M^2$ and $\bar s$ are both large enough.

\section{Summary and Conclusions}

In this work, we have investigated in detail the equatorial motion of spinning test particles around the electrically charged black hole in EiBI gravity. Due to the non-vanishing spin-curvature force, the trajectory of a spinning test particle deviates from the geodesics. For simplicity, we only considered the spin aligned or anti-aligned orbits and explored their properties numerically based on the MPD equations. We have set the parameter $\lambda=1$ and the corresponding cosmological constant $\Lambda=0$ for the electrically charged black hole in EiBI gravity that we considered, and the deviations between the EiBI gravity and GR are controlled by the deviation parameter $\kappa$.

We obtained the effective potential of the spinning test particle with different values of parameter $\kappa$ and fixed the energy of the spinning test particle and confined the spinning test particle moves along the bound orbits. We found that, as $\kappa$ increases, the periastron increases and the apastron decreases, which decrease the orbital eccentricity and narrow the allowed radial range for the spinning test particle. We obtained the relations between the parameter $\kappa$ and velocities $(\dot{r}, \dot{\phi})$ of the spinning test particle, and we found that, as $\kappa$ increases, the radial velocity $\dot{r}$ decreases and the angular velocity $\dot{\phi}$ increases.

{Based on our results for the deviations of orbits induced by the non-zero parameter $\kappa$, we also estimated the possible magnitude of the parameter $\kappa$ in terms of orbits of the observed stars around the Sagittarius A*. Since the radii of periastrons for the observed orbits around Sagittarius A* are at least of the order of $10^3M$, we ignored the effects induced by the spin of the test particle. By naively assuming the observation error as the maximal deviations of the orbits, we found that the parameter $\kappa$ is in the magnitude of $10^{29}$ at least. That is to say, the orbits with large radii can not give a stringent constraint on the parameter $\kappa$ and the investigations of the orbits in the vicinity of the black hole are necessary.}

{In the vicinity of black hole, the effects induced by the spin-curvature force and non-zero $\kappa$ will be significant, for which how the properties of orbits affected by them will be critical to obtain the possible more accurate constraints on the parameter $\kappa$. To obtain the highly dependent relations between the parameter $\kappa$ and orbit parameters, we mainly focused on the orbits in the vicinity of the black hole.} We numerically scanned the causality in the parameter space $(\bar l, \bar s)$, and compared it between the EiBI black hole and the RN black hole. It was found that, for large deviation parameter $\kappa$, two new green colored regions marked V may emerge, where the particle has two stable circular orbits with one subluminal and the other superluminal. The behavior of ISCO parameters, including the radius $r_\text{ISCO}$, orbital angular momentum $\bar{l}_\text{ISCO}$ and energy $\bar{e}_\text{ISCO}$, for different values of spin $\bar s$ and deviation parameter $\kappa$ were studied as well. Our analysis reveled that the spin $\bar s$ dependences of $r_\text{ISCO}$, $\bar{l}_\text{ISCO}$ and $\bar{e}_\text{ISCO}$ have similar behavior to that of RN black hole, namely, they decrease with $\bar s$. Phenomenologically, it means that the spinning particles have a smaller inner edge of accretion disk. For their $\kappa$ dependencies, they decrease monotonously with $\kappa$ when the spin is small, however they changes non-monotonously with $\kappa$ when the spin is large enough.

{By considering the constraint of $\kappa$ from the existence of neutron stars \cite{Pani2011}, the inner edge of accretion disk around a massive black hole ($M/M_\text{sun}>10^2$) in EiBI gravity must be smaller than that around the RN black hole. However, the inner edge of accretion disk around a stellar-mass black hole may greater than that around the RN black hole if both $\kappa/M^2$ and $\bar s$ are large enough. These behaviors will change the dynamics of inspiralling process of the small star in the EMRI system and the status of motion of the matter in the accretion disk. With the improvement of the accuracy of future space gravitational wave detection and black hole shadow observation, we believe that our results will be useful to dig out the nature of the massive black hole ($M/M_{\text{sun}}>10^2$) in EiBI gravity.}

\section*{ACKNOWLEDGMENTS}

This work was supported by the National Natural Science Foundation of China under Grants Nos. 12005174, 12105126, 11947025 and 12165013. K. Yang acknowledges the support of Natural Science Foundation of Chongqing, China under Grant No. cstc2020jcyj-msxmX0370 and ``Fundamental Research Funds for the Central Universities" under Grant No. XDJK2019C051. Y.-P. Zhang acknowledges the support of the Fundamental Research Funds for the Central Universities (Grants Nos. lzujbky-2021-pd08 and lzujbky-2019-ct06), the China Postdoctoral Science Foundation (Grant No. 2021M701531), and ``Lanzhou City's scientific research funding subsidy to Lanzhou University''.



\end{document}